\documentstyle[preprint,aps,twoside,epsf]{revtex}
\begin{document}

\draft

\title{A novel superconducting glass state in disordered
thin films in Clogston limit}

\author{Fei Zhou}

\address{Department of Physics, Princeton University,
Princeton, NJ 08544, USA}

\maketitle

\begin{abstract}
A theory of mesoscopic fluctuations in 
disordered thin superconducting films in a parallel magnetic field
is developed.
At zero temperature and at a sufficiently strong magnetic field,
the superconducting state 
undergoes a phase transition into a state characterized by 
superfluid densities of random signs, instead of a spin polarized
disordered Fermi liquid phase. Consequently, in this
regime, random
supercurrents are spontaneously created in the ground state
of the system, which 
belongs to the same universality class as the two dimensional $XY$ spin
glass.
As the magnetic field increases further, mesoscopic pairing states are 
nucleated in an otherwise homogeneous spin polarized disordered 
Fermi liquid. The statistics of these pairing states is universal
depending on the sheet conductance of the 2D film. 
\end{abstract}

\narrowtext
\section{Introduction}

Recent experiments on thin superconducting films in a parallel magnetic field
\cite{Wu95} have rekindled interest in this field. 
If the thickness of the films $d$ is small enough,
the orbital effect of the magnetic field can be neglected and the 
suppression of 
superconductivity in the film is due to the Zeeman effect
\cite{Abr,Ch62,Clo62}. 
It has been observed that the resistance of such films at low 
temperatures and high enough  magnetic fields exhibits very
slow 
relaxation in time \cite{Wu95}. This 
behavior is characteristic for spin and superconducting glasses.
Below we discuss a possibility that 
mesoscopic fluctuations of superconducting parameters in 
disordered films account for such a behavior.

Mesoscopic physics in a noninteracting electron system has been 
known for a while\cite{Lee,Altshuler85}.
The energy spectrum in a mesoscopic sample was shown to exhibit
Wigner-Dyson
statistics, which is universal, only dependent on the symmetry
of the Hamiltonian\cite{Altshuler86}. 
The long range level repulsion in the energy spectrum
leads to a suppression of fluctuations of levels within 
an energy band of width $E_c=D/L^2$(Thouless
energy). $L$ is the length of the sample,
$D={v_{F}l}/{3}$ is the diffusion constant of the film.
$v_F$ is the Fermi velocity, $l$ is the elastic mean free path.
For an open sample, the fluctuation of number of levels $\delta N$ 
within the energy band of Thouless energy $E_c$ is of order unity, 

\begin{equation}
\frac{<(\delta N)^2>}{<N>^2}=\frac{\beta}{G^2}
\end{equation}  
for a $2D$ film, where the corresponding average number of levels 
$<N>=L^2 d\nu_0E_c =G$, with $\nu_0$ being the average density of states in 
the metal on the Fermi surface. 
$\beta$ is a factor of order unity depending on the symmetry of
the Hamiltonian. 
$G=k_F^2 d l$, is the dimensionless
conductance of the $2D$ normal metal in units
of ${e^2}/{\hbar}$, $k_F$ is Fermi wave length and the 
brackets $\langle\rangle$ denote averaging over realizations of random
potential. Consequently,
the transport is governed by UCF (universal conductance fluctuation) theory. 
The conductance exhibits sample specific
fluctuations, with amplitude $e^2/\hbar$, independent
of the average conductance of the sample\cite{Lee,Altshuler85}.
More
generally, any physical quantity in a mesoscopic sample
consist of an ensemble average part and a sample
specific part due to quantum interference.

On the other hand, disordered superconductors have been studied long ago
\cite{Abr}.
It was shown that
the ground state condensate wave function is 
homogeneous and the critical temperature
remains unchanged in the presence of weak 
nonmagnetic disorders.
To derive the dirty superconductor theory, one has to assume
that 1). the effective interaction constant in the Cooperon channel 
remains the same as in a clean superconductor; 2). the condensate
wave function is translationally invariant; 3). 
the time reversal symmetry is preserved.
The first assumption, though is not true in the thin film limit
where the Coulomb interaction in Cooperon channel can be greatly 
enhanced, is valid in the bulk limit\cite{Finkelshtein}. 
We will assume its validity because
it does not affect the result present in this paper as far as the
renormalized interaction constant is still negative. 
The translation invariance is not a generic symmetry of the original
Hamiltonian in the presence of disorder and the second assumption
is true only after  
the impurity average is taken in the semiclassical limit.
The sample specific quantum interference effect which is of the same
origin of Wigner-Dyson statistics was not taken into account. 
The consequency of such an effect
which breaks the translation invariance is one of the subjects of this article.

The most fundamental aspect of Anderson theory for a dirty
superconductor is the 
{\em absence of spontaneous time reversal symmetry breaking}; that is, the 
stability of a BCS state with respect to possible frustrations,
even when $l$ is much shorter than the coherence
length. This is in contrast to how a ferromagnet responses to
impurities: A spin glass phase which does not have a conventional
long range order but does have Edward-Anderson type long range
order always takes over when impurities are added into the system\cite{Edward}. 
This central issue will be addressed in this article,
in connection with nodes in the spatial dependence of exchange
interactions and the distribution function of the exchange interactions.

Unlike in the noninteracting metal where the mesoscopic
physics is relevant only in a finite sample smaller than the dephasing
length, in the presence of off-diagonal long range order, it 
reveals itself in the thermodynamic limit. 
However, when the elastic mean free path $l$ exceeds
the Fermi wave length $\hbar/k_{F}$, mesoscopic fluctuations of
various physical parameters 
of superconductors are smaller than their
averages \cite{Alt,ZyS,Spivak91,bin}. 
Thus, it seems that they hardly affect macroscopic observable quantities.
It was realized later that
there are situations where mesoscopic fluctuations determine 
macroscopic properties of a superconducting sample. One example is
a superconductor in a magnetic field close to the upper critical field
$H_{c2}$, where the magnetic field dependence of the 
superconducting critical temperature is determined by the mesoscopic 
fluctuations \cite{Spivak95}.
In general, the mesoscopic effects are not only relevant
in a disordered superconductor but also
determinant to the global phase rigidity.

In this paper we consider the case, where the magnetic field is parallel
to the thin superconducting film and the main contribution to the 
suppression of superconductivity by the magnetic field is due to Zeeman 
splitting of electron spin energy levels. 
We show that at low temperatures $T$ and high enough magnetic fields $H$, 
parallel to the film, the 
system exhibits a transition into a state where the local superfluid density 
$N_{s}({\bf r})$ (which is the ratio between the supercurrent density
${\bf J}_{s}$ 
and the superfluid velocity ${\bf V}_{s}$) has a random sign. In this case
the 
system belongs to the same universality class as the two-dimensional XY
spin glass model with exchange interaction of random signs.
We also find that as the magnetic field is decreased from 
above the critical field, mesoscopic pairing states are nucleated
in an otherwise spin polarized disordered Fermi liquid. The characteristic 
length scale at which pairing takes place increases as the critical
field is approached. The statistics of these pairing state 
is universal depending on
the sheet conductance only.

The idea that the superfluid density can be of random signs has a long history 
\cite{Ar,Alt,ZyS,Spivak91,Bul,Ki,KS}. However, in the absence of magnetic 
fields and at zero temperature in 
disordered superconductors ($\xi_{0}\gg l\gg {\hbar}/{k_{F}}$)
the variance of 
the superfluid density, averaged over the superconducting coherence length
$\xi_{0}=\sqrt{{D}/{\Delta_0}}$, turns out to be much smaller than
its
average \cite{Alt,ZyS,Spivak91,bin}
\begin{equation}
\frac{\langle(\delta N_{s})^2\rangle}{\langle
N_{s}\rangle^2}=\frac{1}{G^2}\ll 1,
\end{equation}
$\Delta_{0}$ is the value of the order
parameter at $T,H=0$.
Similarity between Eqs.1,2 suggests the intimate relationship 
between the fluctuations of the superfluid densities 
and universal Wigner-Dyson statistics. 
In fact, Eq.2. follows as a consequency
of the fluctuation of number of levels within Thouless energy band in
a volume of size of the coherence length $\xi_0$.  
As long as $k_{F}l\gg \hbar$, the regions where the 
superfluid density is negative are rare and do not contribute 
significantly to macroscopic properties of superconductors.  
The situation in the presence of a magnetic field 
parallel to the film is different, because the average 
superfluid density decays with $H$ faster 
than its variance.
Hence, at high enough magnetic field the amplitude of the mesoscopic 
fluctuations of $N_{s}({\bf r})$ becomes larger than the average, and the
respective 
probabilities of having positive and negative signs of $N_{s}({\bf r})$
are of the same order even at $k_Fl/\hbar \gg 1$(See below). 
This was first pointed out in an early paper by the author\cite{zhou98b}.

In section 2, we present the qualitative picture of this phenomenon,
emphasising on the sensitivity of mesoscopic fluctuations
of spin polarization energy to the change of the pair potential.
In section 3, we study the mesoscopic fluctuations of the superconducting 
order parameter near the critical regime and show that there are
spontaneously created currents in the ground state.
In section 4, we derive the distribution function 
of the ground state condensate wave function at a magnetic field
higher than the critical one. In section 5,  
we discuss the role of the exchange interaction. 
In section 6, we discuss the mesoscopic effects in
a finite size superconductor.
In conclusion, we propose possible experiments to observe these effects
and point out a few open questions, including
implications on d-wave superconductors.

\section{Qualitative Picture}

A theory of magnetic field induced phase transition which does not 
take into account mesoscopic fluctuations predicts 
\cite{Werthamer66,Maki,Abr} that at low temperatures the superconductor-normal 
metal transition, is of first or second order
depending on whether the parameter $\Delta_{0}\tau_{so}$ is larger or smaller 
than unity respectively. Here, $\tau_{so}$ is the spin-orbit relaxation
time. The $H$-dependence of the order parameter for the two 
limiting cases was discussed extensively in\cite{Werthamer66,Maki}.

From now on, we restrict ourselves to the limit 
$\Delta_{0}\tau_{so}\ll 1$,
where the theory predicts a second order phase transition
between the superconducting state and the normal state.
At $T=0$ and within an approximation which neglects mesoscopic 
effects, the value of the critical magnetic field $H^{0}_{c}$ is the
result of the 
competition between the average superconducting condensation energy density
 $\langle E_{c}\rangle\sim \nu_0 \Delta_0^{2}$ and the
polarization energy of the electron gas in the 
magnetic field.
The average spin polarization energy
density of nonsuperconducting 
electron gas is of order $\langle E_{p}(0)\rangle\sim \nu_{0}(\mu_{B}
H)^{2}$. Its 
relative change in the superconducting state is of order  
$\langle E_{p}(0)\rangle -\langle E_{p}(\Delta)\rangle\sim
\frac{3}{4}\pi\Delta_0\tau_{so}\langle E_{p}(0)\rangle\ll 
\langle E_{p}(0)\rangle$  \cite{Ferrell,An,Ab62}.
 As a result we get an expression 
for the critical magnetic 
field $H^{0}_{c}=H_{cc}(\Delta_0\tau_{so})^{-\frac{1}{2}}\gg 
H_{cc}$. Here $H_{cc}={\Delta_0}/{\mu_B}$ is the
Chandrasekar-Clogston critical magnetic field of the superconductor-normal
metal transition for
$\Delta_{0}\tau_{so}\rightarrow\infty$ and $\mu_{B}$ is the Bohr
magneton.

Consider the mesoscopic fluctuations of the quantities, 
discussed above, in a volume whose size is of the order of the coherence length 
$\xi_{0}$. To calculate the amplitude of mesoscopic 
fluctuations of the polarization energy $\delta E_{p}$, 
we use the conventional diagram technique for
averaging over realizations of random potentials\cite{Abrikosov62}. 
By evaluating the diagrams in Fig.2.d (see Appendix A), we have

\begin{equation}
\sqrt{\langle(\delta E_{p})^{2}\rangle}\propto \frac{1}{G}(\Delta_0\tau_{so}) 
\langle E_{p}(0)\rangle. 
\end{equation}
This part of the polarization energy is sensitive to the change of 
the pair potential($H \sim H_c$) just as the quantum interference effect is 
sensitive to the change of impurity potentials. 
The $\Delta$-dependent part of the mesoscopic fluctuation of
spin polarization energy
can be obtained by calculating the diagrams in Fig.2.d,

\begin{eqnarray}
\sqrt{\langle(\delta E_{p}(\delta\Delta)-\delta E_{p}(\Delta
=0))^{2}\rangle}\propto
\frac{1}{G}(\tau_{so}\Delta_0)
<E_p(0)>\frac{\delta\Delta^2}{\Delta_0^2}.
\end{eqnarray}
Linear in $\delta\Delta$ term vanishes because 
{\em a quasiparticle is reflected into a quasi hole when
it is scattered by $\delta\Delta$}.
As a result, the change of mesoscopic fluctuations of the
spin polarization energy associated with the change of the 
pair potential $\delta \Delta$ of order $\Delta(H)$ is given as

\begin{eqnarray}
\sqrt{\langle(\delta E_{p}(\Delta(H))-\delta E_{p}(\Delta
=0))^{2}\rangle}\propto \sqrt{\langle (\delta
E_{p}(0))^2\rangle}\left(\frac{\Delta(H)}{\Delta_{0}}\right)^{2}.
\end{eqnarray}

Eq.5 shows that below the critical field, though the average cost of 
condensation energy and kinetic energy 
to have a configuration with order parameter equal to zero in
some regions are positive, the mesoscopic
fluctuations of polarization energy associated with such a
configuration are of random signs. 
Since both the polarization energy and the condensation energy are 
fluctuating quantities, $\Delta({\bf r})$ should also be
spatially fluctuating.
Particularly, when $H^0_c$ is approached,
the average cost of energy vanishes, 
the spatial structure of the order parameter
is completely determined by mesoscopic fluctuations
of spin polarization energy.
One elaborates the following argument to confirm this picture.

Consider a domain of size $L_D \gg \xi_{0}$ where the value of 
$\Delta({\bf r})$ differs from its bulk value by a factor of order 
unity. 
An estimate for the energy of such a domain consists of three terms,
namely 
\begin{equation}
\delta E(\Delta)= 
\nu_0 \Delta(H)^{2}d (C_{1}\frac{1}{G}L_D \xi_0 +
C_{2}\frac{H_c^0-H}{H^0_c} L_D^{2} 
+C_{3} \xi_0^2 )
\end{equation}
where $C_{1},C_{2},C_{3}$ are factors of 
order of unity. The first term in Eq.6 corresponds to
the $\Delta$-dependence of mesoscopic fluctuations of polarization
energy and has a random sign. When estimating this term we have taken into
account that regions of size $\xi_{0}$ make independent random
contributions 
into Eq.6. The second and third term are the average condensation
energy and surface (gradient) energy of the domain, respectively.
It follows from Eq.6 that when $L_D \sim
\xi(H)=\xi_{0}{\Delta_0}/{\langle\Delta(H)\rangle}$, there is an
interval of
magnetic fields near the critical one 

\begin{equation}
H_c^0 -H \sim \frac{H^0_{c}}{G^2}
\end{equation}
where the first term is larger than 
the second and the third ones. 
Here 
$\langle\Delta(H)\rangle=\Delta_{0}\sqrt{(H^0_{c}-H)/H^0_{c}}$ is the 
average superconducting order parameter. 
It means that, in this case the spatial 
distribution of 
$\Delta({\bf r})$ is highly inhomogeneous and the amplitude of
the spatial fluctuations of $\Delta({\bf r})$ is of order of its 
average, while the characteristic size of the domains is of order of 
$L_{D}\sim \xi(H=H_{c}^{0}(1-1/{G^2}))\sim \xi_{0}G$. 
At $H$ far away from $H^0_c$, the typical
mesoscopic fluctuations are smaller than the average contribution
and do not change the most probable configuration; they only
introduce exponentially small concentrations of defects
originating from the statistically rare events. 

The second kind of the instability which is intimately connected with 
this inhomogeneous mesoscopic superconducting state is 
spontaneous creation of long range current.
In other words, the superfluid density,
which is the second derivative of the energy with respect to
superfluid velocity in this region has a random sign and 
is no longer positive defined. To see 
this, one should consider states with finite superfluid velocity 
${\bf V}_{s}=\frac{1}{m}(\nabla\chi + {2e}/{c}{\bf A}$), where
$\chi({\bf r})$ is the
phase of the order parameter, $m$ is the electron mass and
${\bf A}({\bf r})$ is the vector
potential, which has a direction parallel to the
film and $m$ is the electron mass. If ${\bf V}_{s}({\bf r})$ is of the
order of the 
critical velocity, all three terms in Eq.6 are modified by factors of
order of
unity when compared with the case ${\bf V}_{s}=0$. 
The second and the third term in Eq.6 decrease with ${\bf V}_{s}$, while
the first 
term is changed in a random direction. This 
means that at high enough magnetic fields, states with nonvanishing value
of ${\bf V}_{s}({\bf r})$ can 
have lower energy than the states with ${\bf V}_{s}=0$, and that the
system is unstable with respect to the creation 
of supercurrents flowing in random directions. 
In this estimate we neglect the
energy of the magnetic field associated with ${\bf V}_{s}({\bf r})$
in the thin film limit.
Since at each point 
of the system the possible energy gain associated with finite value 
of ${\bf V}_{s}({\bf r})$ is independent of the direction of
${\bf V}_{s}$,
the ground state of the system is highly degenerate and  
belongs to the same universality class as $XY$ spin glass with a random sign 
of the exchange interaction. 

In principle, both the spin polarization energy and the condensate
energy fluctuate from region to region. 
Both are related to the density of states, though
the condensate energy should be determined by 
the solution of the self consistent equation.
For the above argument to be true, we have to assume 
that the spin polarization
energy and the condensate energy fluctuate independently.
The argument can be carried out in a similar fashion even if 
these two are partially correlated, as far as {\em they are not fully
correlated}\cite{Kagan}. 
More serious consideration of the existence of inhomogeneous
state is addressed in term of the self consistent equation in the next
section.

It is important to mention that even in the case of small magnetic fields
in the presence of spin orbit scattering the time reversal symmetry is 
broken and the electron wave functions are complex. Therefore there are
currents  in the 
ground state of the system which have random directions. These
currents
exist even in normal metals. 
Diagrammatic calculation leads to an expression for the
correlation function of the current 
density ${\bf J}({\bf r})$ in a normal metal
induced by a magnetic field ($|{\bf r}-{\bf r'}|\gg {\hbar}/{k_F}$)
as shown in \cite{zhou98b}
 
\begin{equation}
\langle{\bf J}_i({\bf r}) {\bf J}_j({\bf r'})\rangle\approx
\delta_{ij}\frac{e^{2}}{\hbar^4 d^{2}}\tau\tau_{so}(\mu_B
H)^{4}\delta({\bf r}-{\bf r'}).
\end{equation}
Here $\tau={l}/{v_{F}}$ is the elastic
mean free time.
It is important to note, however, that for a given configuration of the
scattering potential
and at a given value of the external field the spatial distribution of
${\bf J}({\bf r})$ is a unique function. This implies that the currents
described by Eq. 8 do not
exhibit features which can be associated with superconducting glass
states. In other words, at small $H$ the superfluid density of the
superconducting state is positive which means
that states with nonzero superfluid velocity have larger
energy than the ground state. The rare regions, where
$N_{s}({\bf r})<0$ and the supercurrents  
exist in the ground state, are 
screened effectively due to the Meissner effect. They do not
affect significantly the macroscopic behavior of a sample.

\section{Condensate wave function: $H \leq H_c^0$} 

Below, we will be interested in supercurrents much larger than those
described by Eq.8. Such currents 
are spontaneously created at strong enough magnetic fields as a result of 
the instability associated with the random sign of superfluid density. 
To evaluate the variance of the superfluid density we consider the Gorkov 
equation for $\Delta({\bf r})$\cite{Abrikosov62},

\begin{equation}
\Delta({\bf r})=g\int d{\bf r'}K({\bf r},{\bf r'};
H,{\bf A}({\bf r}),\{\Delta({\bf r})\})
\Delta({\bf r'}),
\end{equation}
where $g$ is the dimensionless interaction constant,  
\begin{eqnarray}
&&K({\bf r},{\bf r'}; H, {\bf A}({\bf r}), \{\Delta({\bf r})\})
\nonumber \\
&&=\nu_0^{-1} k T\sum_\epsilon
G^{\alpha_1\alpha_2}_{\epsilon}({\bf r},{\bf r'}; H,{\bf A}({\bf r}),
\{\Delta({\bf r})\})
\sigma_y^{\alpha_2\alpha_3}
\tilde{G}^{\alpha_3\alpha_4}_{-\epsilon}({\bf r},{\bf r'};
H,{\bf A}({\bf r}), 0)
\sigma_y^{\alpha_4\alpha_1},
\end{eqnarray}
$G^{\alpha \beta}_{\epsilon}({\bf r}, {\bf r}'; H, {\bf A}({\bf r}),
\Delta({\bf r}))$ is
the exact one particle electron Matsubara Green function in
the presence of pair potential $\Delta({\bf r})$, $\alpha$ and $\beta$ 
are spin indexes, $\sigma_{y}^{\alpha \beta}$ is the $y$
component of Pauli matrix and
$\epsilon=(2n+1)\pi kT$ is the Matsubara frequency.
Both
$\Delta({\bf r})$ and $K({\bf r},{\bf r'})$ in Eqs.9 and 10 are
random functions of the realizations of scattering potential in the sample.
Averaging Eq.10 over realizations of the random potential and using the 
approximation $\langle\Delta({\bf r}) K({\bf r},{\bf r'})\rangle=
\langle\Delta({\bf r}; H)\rangle\langle K({\bf r},{\bf r'}, H)\rangle$ we
get the above mentioned expression for $H^{0}_{c}$.

In the case of strong magnetic fields, when 
$\Delta(H,{\bf r})\ll \Delta_0$, we can 
expand Eq.10 in terms of $\Delta({\bf r})$. Since $\Delta({\bf r})$ varies 
slowly over distances of the order of $\xi_0$,
while $\langle K({\bf r},{\bf r'})\rangle$ decays 
exponentially for
$|{\bf r}-{\bf r'}|\gg \xi_0$, we can also make the gradient expansion of 
Eq.10. As a result we get from Eq.10 
\begin{equation}
\left(\xi_{0}^{2}(\nabla-i\frac{2e}{c}{\bf A})^2 + \frac{H^0_c
-H}{H^0_c}\Delta({\bf r})\right)+ 
\int \delta K^0({\bf r}, {\bf r'},H,{\bf A})\Delta({\bf r'}) d{\bf r'}
=\frac{\Delta^3({\bf r})}{2\Delta_0},
\end{equation}
where
\begin{equation}
\delta K^0({\bf r}, {\bf r'})=K^0({\bf r}, {\bf r'}) - 
\langle K^0({\bf r}, {\bf r'})\rangle
\end{equation}
and $K^{0}({\bf r},{\bf r'})=K(\Delta({\bf r})=0,{\bf r},{\bf r'})$.  The 
difference between Eq.11 and the conventional Ginsburg-Landau equation
is the third term in Eq.11 which accounts for mesoscopic fluctuations
of the kernel $K^{0}({\bf r},{\bf r'})$. It is precisely this term, which
at high magnetic fields leads to the random sign
of superfluid density.
To proceed further,  we calculate the correlation function 

\begin{equation}
{\cal C}({\bf r}_1, {\bf r}'_1; {\bf r}_2, {\bf r}'_2)
=\langle\delta K^{0}({\bf r_1},{\bf r'_1}) \delta
K^{0}({\bf r_2},{\bf r'_2})\rangle 
\end{equation}
using the diagrams shown in Fig.2f. And we obtain

\begin{eqnarray}
&&{\cal C}({\bf r}_1, {\bf r}'_1; {\bf r}_2, {\bf r}'_2)
= (2\pi T)^2 \frac{1}{\nu_0^2} 
\sum \exp(i{\bf q} \cdot {\bf r}_1 -i{\bf q}_1 \cdot {\bf r}_2 +
i{\bf q^{'}}_1 \cdot {\bf r}^{'}_2 -i{\bf q}' {\bf r}'_1)
\nonumber \\
&&\delta({\bf q} +{\bf q}'_1 -{\bf q}' -{\bf q}_1)
\delta({\bf Q}_1-{\bf Q}_2+{\bf q}-{\bf q}')
\nonumber \\
&&\sum_{\epsilon, \epsilon'}
(D{\bf q}_1 \cdot {\bf q} +2|\epsilon_1 +\epsilon_2|)
(D{\bf q}' \cdot {\bf q}'_1 +2|\epsilon_1 +\epsilon_2|)
\sigma_y^{\alpha_1-\alpha_1}\sigma_y^{\beta_1-\beta_1}
\sigma_y^{\beta_2-\beta_2}\sigma_y^{\alpha_2-\alpha_2}
\nonumber \\
&& [ C^{\alpha_1\gamma_1}_{2\epsilon}({\bf q})
C^{\beta_1-\gamma_1}_{2\epsilon}({\bf q'})
C^{\gamma_2\alpha_2}_{2\epsilon'}({\bf q}_1)
C^{-\gamma_2\beta_2}_{2\epsilon'}({\bf q}''_1)
C^{\gamma_1\gamma_2}_{\epsilon+\epsilon'}({\bf Q}_1)
C^{\gamma_1\gamma_2}_{\epsilon+\epsilon'}({\bf Q}_2)
\nonumber \\
&&+C^{\alpha_1\gamma_1}_{2\epsilon}({\bf q})
C^{\beta_1\gamma_1}_{2\epsilon}({\bf q'})
C^{\gamma_2\alpha_2}_{2\epsilon'}({\bf q}_1)
C^{\gamma_2\beta_2}_{2\epsilon'}({\bf q}''_1)
D^{\gamma_1\gamma_2}_{\epsilon+\epsilon'}({\bf Q}_1)
D^{\gamma_1\gamma_2}_{\epsilon+\epsilon'}({\bf Q}_2) ].
\end{eqnarray}
The large distance asymptotics of the correlation function
in Eq. 14 takes the form, 

\begin{eqnarray}
&&{\cal C}({\bf r}_1, {\bf r}'_1; {\bf r}_2, {\bf r}'_2)
=\frac{1}{16 G^2}
\left[\xi_0^2\delta({\bf r}_1-{\bf r}_2)\delta({\bf r}'_1 -{\bf r}'_2) 
\delta({\bf r}_1-{\bf r}'_1)
+\delta({\bf r}_1-{\bf r}'_2)\delta({\bf r}'_1 -{\bf r}_2) 
\frac{\xi^4_0}{|{\bf r}_1-{\bf r}'_1|^4}\right].
\end{eqnarray}

Eq.15 characterizes the sample specific interference 
effect on the Cooperon propagator defined in Eq. 10. 
It determines the mesoscopic fluctuations of
the superconducting order parameter, which represent the deviation of the 
exact ground state from the translationally invariant state. 
Employing the perturbation theory with respect to $\delta 
K^{0}({\bf r},{\bf r'})$ 
we get from Eq.11 an expression for the correlation function 
of the mesoscopic fluctuations of the superconducting order parameter
$\delta\Delta({\bf r}; H)=\Delta({\bf r};H)-\langle\Delta(H)\rangle$

\begin{eqnarray}
&&\langle\delta \Delta({\bf r})
\delta \Delta({\bf r'})\rangle
=\Delta^2(H)
\int d{\bf r}_1
d{\bf r}'_1
d{\bf r}_2
d{\bf r}'_2
L({\bf r}, {\bf r}_1)L({\bf r}', {\bf r}'_1)
{\cal C}({\bf r}_1, {\bf r}'_1; {\bf r}_2, {\bf r}'_2)
\nonumber \\
&&L({\bf r}, {\bf r}_1)=
\sum_{q}\frac{\exp(i {\bf q} ({\bf r} -{\bf r}_1))}
{\xi_0^2 {\bf q}^2 + (H-H_c^0)/H_c^0} 
\end{eqnarray}
Taking into account Eq.15, 
\begin{eqnarray}
\langle\delta \Delta({\bf r})
\delta \Delta({\bf r'})\rangle
\propto
\frac{\Delta^2_0}{G^{2}} \left \{ \begin{array}{cc}
1- \frac{1}{2}\ln(\xi(H)/\xi_0)[|{\bf r} -{\bf r'}|/\xi(H)]^2, 
\mbox{$|{\bf r} -{\bf r'}| \ll 
\xi(H)$}; \\
\exp[-|{\bf r} -{\bf r'}|/\xi(H)], \mbox{$|{\bf r} -{\bf r'}| \gg
\xi(H)$}
\end{array}\right.
\end{eqnarray}
It follows from Eq.17 that the
amplitude of the fluctuations of the order parameter in the
two-dimensional
case is almost independent of $H$, but the average order parameter
decreases with $H$. As a result, perturbation theory holds
as long as
${\langle\Delta(H)\rangle}/{\Delta_0}=\sqrt{(H^{0}_{c}-H)/H^{0}_{c}}\gg
G^{-1}$.
This also justifies Anderson theory of dirty superconductors in
the absence of an external magnetic field:
the ground state is approximately 
translationally invariant though the translation 
invariance is not a generic symmetry of original Hamiltonian in
the presence of impurity potentials.

Eq.17 implies that a homogeneous superconducting state becomes
unstable against the mesoscopic fluctuations near the critical
point. 
Such an instability against the inhomogeneous state
can also be visualized if  
the magnetic field is decreased from above the critical field.
The generalized curvature characterizing the stability  
of a metal or at $\Delta({\bf r})=0$, is defined as, 

\begin{eqnarray}
{\cal O}({\bf r}, {\bf r}')=\frac{\delta^2 E(\{\Delta({\bf r})\})}
{\delta \Delta({\bf r})\delta \Delta({\bf r}')}
\end{eqnarray}
where $E(\{\Delta\})$ is the energy of a configuration $\{\Delta({\bf r})\}$,
and $\delta/\delta \Delta({\bf r})$  stands a functional derivative. 
Following Eq. 11, we obtain
\begin{eqnarray}
{\cal O}({\bf r}, {\bf r}')=-(\xi_0^2(\nabla - i\frac{2e}{c}{\bf A})^2
+\frac{H_c^0 -H}{H_c^0})\delta({\bf r}-{\bf r}')
-\delta K^0({\bf r}, {\bf r}')
\end{eqnarray}
Eq.19 shows that 

\begin{equation}
<({\cal O}({\bf r}_1, {\bf r}'_1)
-<{\cal O}({\bf r}_1, {\bf r}'_1)>)
({\cal O}({\bf r}_2, {\bf r}'_2)-
<{\cal O}({\bf r}_2, {\bf r}'_2)>)>
={\cal C}({\bf r}_1, {\bf r}'_1; {\bf r}_2, {\bf r}'_2).
\end{equation}
Generally speaking, the curvature matrix in Eq. 18 is not a positive defined 
one because of fluctuations. However, well above
the critical point, the probability to find the region where
the curvature is negative is exponentially small; the
ground state will be a normal metal with exponentially small
concentration of superconducting droplets.
The probability to find these droplets will be discussed in detail in
the next section. Here we want to point out that 
when the critical point is approached, the mesoscopic fluctuations 
of the curvature matrix becomes larger than the positive defined
average part(first two terms) and the probability of finding
the superconducting regions in the ground state becomes of order of one. 
This again implies that the most probable configuration near $H_c$
is an inhomogeneous state.

The instability against spontaneous creation of current state
can be demonstrated via studying the superfluid density defined as 
\begin{eqnarray}
{\cal N}^{ij}({\bf r}, {\bf r}')=\frac{\delta^2 E}{\delta V_{si}({\bf r})
\delta V_{sj}({\bf r}')},
\end{eqnarray}
which can be written in term of exact Green functions 
when $\Delta(H) \ll \Delta_0$

\begin{eqnarray}
{\cal N}^{ij}({\bf r}, {\bf r}')={\bf e}_i{\bf e}_j\frac{e^2kT}{m}
\sum_{\omega}\int d{\bf r}_3 d{\bf r}_4
\nabla_{\bf r}
G^{\alpha_1\alpha_2}_\omega({\bf r}, {\bf r}_3)\Delta({\bf r}_3)
\sigma_y^{\alpha_2-\alpha_2}
\nonumber \\
G^{-\alpha_2\alpha_3}_{-\omega}({\bf r}_3, {\bf r}')
\nabla_{{\bf r}'}
G^{\alpha_3\alpha_4}_\omega({\bf r}', {\bf r}_4)\Delta({\bf r}_4)
\sigma_y^{\alpha_4-\alpha_4}
G^{-\alpha_4\alpha_1}_{-\omega}({\bf r}_4, {\bf r}).
\end{eqnarray}
Expanding Eq.22 in terms of
$\delta \Delta({\bf r},H)\ll \Delta(H)$
we get an expression for the nonlocal superfluid density 
${\cal N}^{ij}({\bf r},{\bf r'})$, which is valid as long as 
${\langle\Delta(H)\rangle}/{\Delta_{0}}\gg G^{-1}$ and $\delta
N_{s}=(N_{s}-\langle N_{s}\rangle)\ll \langle N_{s}\rangle$

\begin{equation}
{\cal N}^{ij}({\bf r}, {\bf r}')=<{\cal N}^{ij}_s({\bf r}, {\bf r'})>+
\delta {\cal N}^{ij}_s({\bf r}, {\bf r}').
\end{equation}
The average superfluid density is $\delta$-correlated over distances
larger than $l$, 
\begin{equation}
\langle {\cal N}^{ij}_{s}({\bf r},{\bf r'})\rangle=
N_s^0\frac{\Delta(H)^2}{\Delta_0^2}
\delta_{ij}\delta({\bf r}-{\bf r'}),
\end{equation}
where $N_{s}^{0}=eN({l}/{\xi_0})^2$ is the average superfluid density
at $H=0$ and $N$ is the electron concentration in the metal.

$\delta {\cal N}_{ij}({\bf r}, {\bf r}')$ is determined by the diagrams in
Fig.2g,2h.
Following the calculations in Appendix B, 

\begin{eqnarray}
\frac{\langle\delta {\cal N}^{ij}_s({\bf r}_1, {\bf r}_1') 
\delta {\cal N}^{i'j'}_s({\bf r}_2, {\bf r}_2')\rangle}{(N_s(H))^2}
= \frac{ \langle \delta \Delta({\bf r}_1)
\delta \Delta({\bf r}_2)\rangle^2}{\langle\Delta(H)\rangle^{4}}
\delta_{ij}\delta_{i'j'}\delta({\bf r}_1 - {\bf r}_1')
\delta({\bf r}_{2} - {\bf r}_2')
\nonumber \\+
\frac{1}{16 G^2} 
\frac{\xi_0^4}{|{\bf r}_{1} -{\bf r_1'}|^4}
(\delta_{ii'}\delta_{j'j'}\delta({\bf r}_{1} - {\bf r}_{2})
\delta ({\bf r_1'} - {\bf r}_2')
+\delta_{ij'}\delta_{i'j}\delta({\bf r}_{1} - {\bf r'}_{2})
\delta ({\bf r_1'} - {\bf r_2})).
\end{eqnarray}
Here $N_s(H)=N_s^0(\Delta(H)/\Delta_0)^2$ is the superfluid density at $H$.
The first term in Eq.25 is connected to the fluctuations of the order 
parameter $\Delta({\bf r})$ in Eq.16 (the corresponding diagram is shown
in Fig.2g).
The second term in Eq.25 is related to the fluctuations of the Green
functions $G_{\omega}({\bf r},{\bf r}')$
(the corresponding diagram is shown in Fig.2h).
When the magnetic field is close to the critical one, i.e.
$|H-H_{c}^{0}|/H^{0}_{c}\sim G^{-2}$, the amplitude of fluctuations
of the superfluid 
density becomes of the order of the average {$\delta N_{s}\sim \langle 
N_{s}\rangle$; the 
local value of the superfluid density, averaged over the size $\xi(H)$, 
becomes of random sign and the system is unstable with
respect to spontaneous creation of supercurrents.

When ${|H-H^0_{c}|}/{H^0_{c}} \ll G^{-2}$, one can neglect the second
term in brackets in Eq.11. 
Rescaling ${\bf r}$ and $\Delta({\bf r})$ as
\begin{eqnarray}
{\bf r}=\frac{1}{2}{y} G\xi_0,  
\Delta({\bf r})=\frac{\Delta_0}{2G}f(\frac{{\bf r} -{\bf r}_0}{ G\xi_0}),
\end{eqnarray}
we obtain a dimensionless equation for $f({y})$ which
represents a continuous version of $XY$ spin-glass model

\begin{equation}
\nabla_{{y}}^{2} f({y}) +\int d{y'} \delta k({y},
{y'}) f({y'})=f^3({y}), 
\end{equation}
where $\langle\delta k({y},{y'})\rangle=0$ and the correlation
function
\begin{eqnarray}
&&\langle\delta k({y}_1, {y'}_1) \delta k({y}_2,
{y'}_2)\rangle
=\delta ({y}_1 -{y}_2)\delta
({y'}_1-{y'}_2)\frac{G^{-2}}{({y}_1 - {y'}_1)^4}
+\delta( y_1 - y_2)\delta( y'_1 - y'_2)
\delta( y_1 - y'_1).
\end{eqnarray}
In this limit,
it follows from Eqs.26-27 that the amplitude 
of spatial fluctuation of the 
modulus of the order parameter $\delta\Delta({\bf r})\sim
\langle\Delta(H)\rangle \sim \Delta_0/ G$ is of order of 
its average. The characteristic spatial scale of the fluctuations of
$\delta\Delta({\bf r})$ is of order of $L_D$.
The important feature of Eq.27 is that the sign of the second term in Eq.27
fluctuates randomly which corresponds to the random sign of the superfluid
density.
The spontaneously created supercurrents in this case have random
directions, their typical amplitude is
of order of $J^{sc}_{s}\sim N_{s}^{0}{\hbar}/{G^3\xi_0}$ and their
characteristic scale of spatial correlations is also of order of $L_D$.
The current in Eq.8 is negligible compared with $J^{sc}_{s}$ when
$l/l_{so} \ll G^{-1}$, where $l_{so} \sim \sqrt{D\tau_{so}}$.

The fact that the sign of $N_{s}$ is random is
especially 
clear in the case of a large magnetic field, when
$H-H^0_c \gg H^0_c G^{-2}$. 
In this case, $\Delta({\bf r})$ can be nonzero only due to existence the 
rare regions, where $\delta k({x},{x'})$ is much larger than the
typical
value given in Eq.28. Thus, the spatial dependence of the modulus of 
the order 
parameter has the form of superconducting domains embedded in a normal
metal. 
These
regions are 
connected via the Josephson effect. We calculate the average critical 
current of the junctions
\begin{equation}
\langle J_{c}\rangle \propto
G\frac{e^2}{\hbar}\frac{D}{L_0^2}\exp\left(-\frac{L_0}{\xi_{0}}
\sqrt{\frac{H^0_{c}}{H - H^0_c}}\right)
\end{equation}
which decays exponentially with the average distance between the 
superconducting droplets $L_0$.
On the other hand, the amplitude of fluctuations of $J_c$ 
 decays only as a power of $L_0$.
\begin{equation}
\langle(\delta J_{c})^{2}\rangle\propto
\left(\frac{e^2}{\hbar}\frac{D}{L^2_0}\right)^2. 
\end{equation}
 As a result, the amplitude 
of the fluctuations in this regime turns out to be larger than the
average, hence $J_{c}$ has a random sign. 
As argued in section 5, such a distribution function of 
exchange interaction
is a generic one when the spins are polarized.

We should emphasis that the Josephson coupling in Eq.30 is derived 
in the limit when $L_0$ is much longer than the coherence length
of the superconducting domains. It doesn't depend on the value 
of $\Delta_0$ in the superconducting domains. This is because
the effective transmission coefficient of Cooper pairs over a
distance of $L_0$ is exponentially small at an energy higher than
$D/L^2\ll \Delta_0$.  In contrast, when $L_0 \ll \xi_0$, the effective 
transmission coefficient of Cooper pairs is independent of $\epsilon$
at an energy smaller than $\Delta_0$. In this limit,
$J_c \propto \Delta_0$.

It is well known \cite{vil} that at $T=0$ the long range order of the
ground state of the two-dimensional $XY$ model is destroyed by an 
arbitrary
small
concentration of "antiferromagnetic"  bounds.
As we have mentioned above in the case $H\ll H^{0}_{c}$,  regions where
$N_{s}({\bf r})<0$,
exist with small but finite probability. 
In this case, however, the properties of a superconductor are
different from the $XY$ model because the supercurrents
spontaneously created in these regions are screened
by the Meissner effect. 
Thus at $H,T=0$ a
superconducting film should exhibit the conventional long 
range order. 

{\em This implies that there is a critical magnetic field 
$H_{SG}<H_{c}^{0}$ at $T=0$, where
the system undergoes a phase transition from a superconducting state to a
superconducting 
glass state}. The typical distance of the rare regions is
estimated in the next section as $L_d$(See Eq. 63). At $H_{SG}$,
it should become comparable with the penetration depth, $\lambda(H)=
\lambda_0^2/d\times (H/H_c^0-H)^2$, or 

\begin{equation}
\lambda(H_{SG})\sim \xi(H_{SG})\exp(G^2\frac{H^0_c-H_{SG}}{H^0_c}).
\end{equation}
It yields

\begin{equation}
\frac{H^0_c-H_{SG}}{H^0_c}\sim \frac{1}{G^2}\ln (G^3 
\frac{\lambda^2_0}{d\xi_0}), \end{equation}
$\lambda_0$ is the zero temperature penetration 
depth of a bulk superconductor.
Eqs. 7,32 show that up to a log-factor, the transition
between the superconductor
and the superconducting glass state takes place at the magnetic
field when $\delta \Delta \sim \Delta(H)$.
The interval of magnetic fields where the system is in the 
superconducting glass state is indicated in Fig.1.

The superconducting glass state which arises due to orbital magnetic field 
effects has been considered in numerous 
papers (See for example \cite{Lub,vin,FishG}). The qualitative difference 
between \cite{Lub,vin,FishG} and
above considered cases is that in the latter case the system exhibits 
glassy behavior as soon 
as $H > H_{c1}$ and vortices begin to penetrate the superconductor.  
Furthermore, the state we discussed 
here is also different from Fulde-Ferrell-Larkin-Ovchinnikov
state \cite{Fulde64,Larkin64}, which becomes a metal stable state in dirty
superconductors when $\xi_{0}\gg l$ \cite{asl}.

\section{Optimal Superconducting Droplets: $H \geq H_c^0$}

One of the consequences of the mechanism discussed above
is the {\em instability of the spin polarized disordered Fermi liquid}
well above the critical magnetic field. 
As argued before, though the average curvature of the normal metal state 
(${\cal O}({\bf r}, {\bf r}')$ evaluated at $\Delta=0$) is positive defined,
its mesoscopic fluctuations have 
random signs because of the mesoscopic
fluctuations of the spin polarization energy. 
In the regions where the spin polarization energy cost to form
superconducting pairing state is much lower than the average
energy cost,
the fluctuations of the curvature
are of large negative value comparable to its positive average such that
the normal metal with $\Delta=0$ becomes unstable.  As a result,  
above the critical field $H_c^0$, the superconducting pairing 
correlations are established at mesoscopic scales 
in the different regions in the normal metal and  
couple with each other via exchange interactions of random signs.  
This argument was present in another early paper by the author\cite{zhou98a}.

In this section, 
we study the probability to find regions  
where the superconducting pairing states are formed at 
mesoscopic scales at $H > H_c^0$.  
At high magnetic fields
in the strong spin-orbit scattering limit, the statistics of these pairing 
states can be studied with the help of the generalized 
Landau-Ginsburg equation, 
\begin{equation} 
\left[\xi_{0}^{2}\left(\nabla-i\frac{2e}{c}{\bf A}\right)^2 + \frac{H^0_c
-H}{H^0_c}\right]\Delta({\bf r})+ 
\int \delta K^0({\bf r}, {\bf r}',H)\Delta({\bf r}')d{\bf r}'
=\frac{\Delta^3({\bf r})}{2\Delta^2_0},
\end{equation}
which is valid when $H-H_c^0$ is small compared
with $H_c^0$ and when the spatial variation of the pairing wave
function $\Delta({\bf r})$ over distance $\xi_0$
is negligible. Here $\xi_0=\sqrt{D/\Delta_0}$,
${\bf A}$ is the vector potential of external perpendicular magnetic field.

Eq.33 is a {\em nonlinear} equation in terms of $\Delta({\bf r})$,
with a {\em nonlocal}  
$\delta K^0({\bf r}, {\bf r}')$ potential
originating from the oscillations of the wave functions
of cooper pairs. 
Generally speaking, it is
qualitatively different from the 
Schroedinger equation of an electron in the presence 
of random impurity potentials\cite{Lifshitz,Halperin,Zittartz,Igor}.
These complications arise naturally in the study of the 
interplay between the mesoscopic effects and the superconductivity 
and are the generic features of {\em strongly correlated} mesoscopic
systems. In fact, this nonlocal structure of the potential
in Eq.33 leads to the superconducting glass state. 

At $H-H_c^0 \gg H_c^0/G^2$, the 
optimal configurations which determine the
macroscopic properties of the sample
turn out to be the superconducting droplets embedded inside
the disordered Fermi liquid, with the phases of each 
droplet coupled via random exchange interaction. 
Such a configuration can be characterized by three parameters:
A). the typical size of the droplet, $L_f$;
B). the typical distance between the droplets, $L_d \gg L_f$;
C). the typical value of the order parameter inside each droplet.
In the following, we will discuss the 
statistics of the mesoscopic pairing states in this regime. 
In the leading
order of $(L_f/L_d)^2$ the statistical property of the formation of
superconducting pairing states at mesoscopic scales 
is similar to that of the impurity band tails 
\cite{Lifshitz,Halperin,Zittartz,Igor}.

The calculation of such a 
probability is closely connected to the evaluation of tails of distribution 
functions of mesoscopic fluctuations \cite{AKL,KM,EF}. 
However, in the present case, $\delta K^0({\bf r}, {\bf r'})$
is determined by the fluctuations 
integrated over the whole energy spectrum 
instead of single energy level. Thus, we believe it is of a Gaussian 
form and the statistical property 
of the random potential $\delta K^0({\bf r}, {\bf r})$ 
is determined by its second moment. General case is discussed in
Appendix C.

The pairing wave function of the most probable configurations is 
given as

\begin{equation} 
\Delta({\bf r})=\sum_{\alpha}\Delta_\alpha \eta_\alpha({\bf r}),
\int d{\bf r} \eta_\alpha({\bf r})\eta_\beta({\bf r})\propto 
\delta_{\alpha\beta}.
\end{equation}
Note $\eta({\bf r})$ introduced in this way is dimensionless.
For such a configuration to have lower energy than the normal 
state,  
\begin{equation}
\int d{\bf r} d{\bf r}' \Delta({\bf r}){\cal O}({\bf r}, 
{\bf r}')\Delta({\bf r}') <0,
\end{equation}
where ${\cal O}({\bf r}, {\bf r'})$ is given by Eq. 19.

The total energy of such a configuration consists of cross terms
corresponding to the coupling between
different droplets. 
The coupling between the droplets decays as the distance
increases.
When the size of the droplets is much smaller than the distance 
between them,
the typical magnitude of the coupling
between different droplets is much smaller than that of
the coupling within one droplet.   
We are going to neglect such terms in 
the estimate of the probability of the droplets
in the leading order of $o(L^2_f/L^2_d)$.

Thus, to have $l$ droplets in the normal metal,   
$l$ independent inequalities have to be satisfied

\begin{equation}
\Delta^2_\alpha \left[
\int d{\bf  r}\eta_\alpha({\bf r})\left(\xi_0^2\nabla^2 +\frac{H^0_c-H}{H^0_c}
\right)
\eta_\alpha({\bf r}) + 
\int d{\bf r} d{\bf r}'  \eta_\alpha({\bf r})
\delta K^0({\bf r}, {\bf r}') \eta_\alpha({\bf r}')\right] < 0. 
\end{equation}
(We assume there is no perpendicular magnetic field.) 
Furthermore, we can write down the probability to have
superconducting pairing states at $H \gg H_c^0$ in term of the sum
of probability to have certain number of droplets 

\begin{equation}
{\cal P}(\{\eta(x)\})=\sum_{l} P_l(\{\eta_\alpha\}|\alpha=1,...,l).
\end{equation}
To simplify the notation, we introduce
\begin{equation}
O_{LG}=\xi_0^2 \nabla^2 +\frac{H_c^0-H}{H_c^0},
K_M=\delta K^0({\bf r}, {\bf r}').
\end{equation} 
Taking into account
$D\eta({\bf r})=\Pi_\alpha D\eta_\alpha$,
we have 

\begin{eqnarray}
P_l(\{\eta_\alpha\}|\alpha=1,...,l)
=\int P(\{K_M\}) \Pi_\alpha 
N^l \int \theta(-E_\alpha+F_\alpha) 
D\eta_\alpha DK_M
\label{probdrop}
\end{eqnarray}
where
\begin{eqnarray}
&&E_\alpha(\{\eta({\bf r})\})=
\int d{\bf r} \eta_\alpha({\bf r}) {O_{LG}} \eta_\alpha({\bf r}),
\nonumber \\
&&F_\alpha(\{\eta({\bf r})\})=\int d{\bf r} d{\bf r}' 
\eta({\bf r})_\alpha K_M({\bf r}, {\bf r}') \eta_\alpha({\bf r}'),
\end{eqnarray}
and $N$ is a normalization constant.
$P(\{ K_M\})$ is the distribution function of $K_M$; 
$D\eta, DK_M$ represent functional integrals.
We use the following equality to transform the step function 
into integrals, 

\begin{equation}
\theta(-E_\alpha+F_\alpha)=
\int_{-\infty}^{0} dg_\alpha 
\int_{-\infty}^{+\infty}dh_\alpha 
\exp\left[ih_\alpha \left(E_\alpha-F_\alpha - g_\alpha\right)\right]. 
\end{equation}
Eq. 37 is reduced to
$P_l(\{\eta_\alpha\}|\alpha=1,...,l)
=\Pi_{\alpha}\rho_\alpha$. 
In the Gaussian approximation, the statistics of 
$\delta K^0({\bf r}, {\bf r}')$ is completely determined 
by the second moment of the correlation function, or

\begin{equation}
<\Pi_{i=1}^{2m}\delta K({\bf r}_i, {\bf r}'_i)>
=\Pi_{i=1}^{m}{\cal C}({\bf r}_{2i}, {\bf r}'_{2i}; 
{\bf r}_{2i+1}, {\bf r}'_{2i+1}),
\end{equation}
where 
${\cal C}({\bf r}_1, {\bf r}'_1; {\bf r}_2, {\bf r}'_2)$
is given in Eq.15.

In this case,
$\rho_\alpha$ can be simplified in a closed form as 

\begin{eqnarray}
&&\rho_\alpha
= N \int {\rm erfc}[   
\frac{E_\alpha(\{\eta({\bf r})\})}
{\sqrt{2\sigma_\alpha(\{\eta({\bf r})\})}}] D\eta_\alpha({\bf r})
\nonumber \\
&&\sigma_\alpha={\int d{\bf r}_1 d{\bf r}'_1  \int d{\bf r}_2 d{\bf r}'_2  
{\cal C}({\bf r}_1, {\bf r}'_1; {\bf r}_2, {\bf r}'_2) 
\eta_\alpha({\bf r}_1) \eta_\alpha({\bf r}'_1) 
\eta_\alpha({\bf r}_2) \eta_\alpha({\bf r}'_2) } 
\end{eqnarray}
with ${\rm erfc}({a}/{b})=\int_a\exp(-x^2/2b^2)/\sqrt{2\pi b^2} dx$.
One can evaluate the functional integral $D\eta_\alpha({\bf r})$ in the saddle
point approximation as long as $H-H_c^0 \gg H_c^0/G^2$. 
The saddle point equation of Eq.43 can be obtained by
minimizing the argument of the error function. 
\begin{eqnarray}
&&O_{LG}\eta_\alpha({\bf r})  + {\bf S}{\int d{\bf r}_1d{\bf r}'_1 d{\bf r}'  
{\cal C}({\bf r}, {\bf r}'; {\bf r}_1, {\bf r}'_1) 
\eta_\alpha({\bf r}_1) \eta_\alpha({\bf r}'_1)\eta_\alpha({\bf r}')}=0 
\nonumber\\
&&{\bf S}= \frac{\int d{\bf r} \eta_\alpha({\bf r})O_{LG}\eta_\alpha({\bf r})}
{\int d{\bf r}_1 d{\bf r}'_1 d{\bf r}_2 d{\bf r}'_2
{\cal C}({\bf r}_1, {\bf r}'_1; {\bf r}_2, {\bf r}'_2) \eta_\alpha({\bf
r}_1) 
\eta_\alpha({\bf r}'_1)\eta_\alpha({\bf r}'_2)\eta_\alpha({\bf r}_2)}
\nonumber\\
\label{saddle}    
\end{eqnarray}
The solution of the saddle
point equation $\eta_s({\bf r})$ determines the shape of the optimal droplets. 
To carry out the functional integral of $\eta_\alpha({\bf r})$, one can expand
$\eta({\bf r})$ around the saddle point,
\begin{equation}
\eta({\bf r})=\eta_{s}({\bf r}) + \delta \eta({\bf r}),
\delta \eta({\bf r})=\sum_{n} a_n \eta_n({\bf r})
\end{equation}
where $\eta_n({\bf r})$ are the eigenstates of the operator 
$\Gamma({\bf r}, {\bf r}')$ generated via second functional
derivative of the argument in the error function with respect to
$\eta({\bf r})$ at $\eta({\bf r})=\eta_s({\bf r})$. Our final result 
barely depends on the detailed structure of $\Gamma({\bf r}, {\bf r}')$ 
and we do not give an explicit form here.
Performing the Gaussian integral of $\delta \eta({\bf r})$
around the saddle point, taking into account the normalization condition,
we obtain,

\begin{equation}
\rho_\alpha={\rm erfc}(\frac{E_{s}}{\sqrt{2\sigma_{s}}})
\frac{det' \Gamma({\bf r}, {\bf r}')}
{det\left\langle{\cal O}({\bf r}, {\bf r}')
\right\rangle}
\int [da_0]
\end{equation} 
where
\begin{eqnarray}
&&{E_{s}}= {\int d{\bf r} \eta_s({\bf r})O_{LG}\eta_s({\bf r})}
\nonumber \\
&&{\sigma}_{s}=
{\int d{\bf r}_1 d{\bf r}'_1 d{\bf r}_2 d{\bf r}'_2
{\cal C}({\bf r}_1, {\bf r}'_1; {\bf r}_2, {\bf r}'_2) \eta_s({\bf r}_1) 
\eta_s({\bf r}'_1)\eta_s({\bf r}'_2)\eta_s({\bf r}_2)}
\end{eqnarray}
$E_{s}/{\sqrt{2 \sigma_{s}}}$ is the argument of
error function in Eq.43 evaluated at $\eta({\bf r})=\eta_s({\bf r})$.
$'$ indicates the exclusion of the zero eigenvalue. 
The last integral in Eq.46 corresponds to the contribution from
the zero eigenvalue state, originating from
the translation invariance of the saddle point equation, 
with $2-fold$ degeneracy   
\begin{equation}
\eta_{0i}({\bf r})=\frac{L_0 \partial_{{0i}} \eta_s({\bf r}-{\bf r}_0)}
{\sqrt{\int \eta_s^2 d{\bf r}}},
\end{equation} 
$i=x,y$\cite{Halperin,Zittartz}. 
Here $L_0$ is the characteristic length of the droplets determined via
the normalization condition
\begin{equation}
1=\frac{\int d{\bf r} L_0^2 (\nabla \eta_s)^2}
{2\int \eta^2_s d{\bf r}}.
\end{equation}
Thus,
\begin{equation}
\int [da_0]=\frac{1}{L_0^2}\int_{v_a} dx_\alpha dy_\alpha. 
\end{equation}
The spatial integral is performed only in the region $v_\alpha$ where
no other droplets are present.  Using the following rescaling 

\begin{eqnarray}
&&{\bf r}={y}{L_f},~
\nabla=\nabla_y L_f^{-1},~
\eta_s({{\bf r}})=\eta_s(\frac{y}{L_f}),~
\nonumber \\ &&{\cal C}({\bf r}, {\bf r}';{\bf r}_1, 
{\bf r}'_1)=\frac{1}{G^2} \frac{\xi_0^2}{L_f^6}  
\tilde{\cal C}(y, y'; y_1, y'_1),~
L_f=\xi_0(\frac{H^0_c}{H -H^0_c})^{1/2},
\end{eqnarray}
we can express $E_{s}$, $\sigma_{s}$
in term of dimensionless $\eta_s(y)$ 
 
\begin{eqnarray}
&&{E_{s}}
=B\frac{H-H^0_c}{H^0_c}L_f^2,~
\sigma_{s}=A^2\frac{\xi_0^2 L_f^2}{G^2}
\nonumber \\
&&B=\int dy {\eta_s}(y)(-\nabla_y^{2}+ 1){\eta_s}(y),
A^2={\int dy_1 dy'_1 dy_2 dy'_2
\tilde{\cal C}(y_1, y'_1; y_2, y'_2){\eta_s}(y_1) {\eta_s}(y'_1)
{\eta_s}(y'_2){\eta_s}(y_2)}
\end{eqnarray}
where $B, A^2$ are the dimensionless quantities
of order of unity depending on the details of ${\eta_s(y)}$. 
$\eta_s$ satisfies the dimensionless saddle point equation 
\begin{eqnarray}
(-\nabla^{2}_y+1){\eta_s}(y)
+ \int dy_1dy'_1 dy'  
\tilde{\cal C}(y, y'; y_1, y'_1)
\eta_s(y_1)\eta_s(y'_1)\eta_s(y')=0 
\end{eqnarray}
and at $y=\infty$, $\eta_s(y)=0$. 
If ${\eta_s(y)}$ is a Gaussian function, $B=2$, $A=18/\pi^2$. 
We also estimate 
that 
\begin{equation}
\frac{det\Gamma({\bf r}, {\bf r}')}{det\left\langle{\cal O}({\bf r}, 
{\bf r}') \right\rangle}
\approx 1+o(\frac{1}{G})
\end{equation}.

Collecting all the results, we have 
\begin{equation}
P_l(\{\eta_\alpha\}|\alpha=1,...,l)
= \frac{V^l}{l!}  
\left\{\frac{1}{ L_f^2}
{\rm erfc}\left[\frac{{B G}}{A}\left(\frac{H_c -H}{H_c}\right)^{1/2}\right]
\right\}^{l}
\end{equation}
where $V^l/l!$ is from the spatial integral in Eq.46, excluding the 
overlap between different droplets. We take into account $L_0\sim L_f$.
It is easy to confirm that the average number density of the droplets
is

\begin{equation}
\rho=\frac{1}{V}\frac{\sum_l P_l l}{\sum_l P_l}= \frac{1}{L_f^2}
{\rm erfc}\left[\frac{B G}{A} \left(\frac{H 
-H^0_c}{H^0_c}\right)^{1/2}\right]. \end{equation}

The distribution function of the amplitude of the order parameter
$\Delta$ in a droplet
can be calculated in a similar way. 
In this case, the amplitude $\Delta$ is determined by the
nonlinear term in Eq.33 and the probability to have a superconducting 
droplet with order parameter equal to $\Delta$ is

\begin{equation}
P(\Delta)= \int \frac{{2\Delta N_\alpha}}{\Delta^2_0}
P({K_M})\delta(N_\alpha (\frac{\Delta}{\Delta_0})^2 +E_\alpha -F_\alpha)
dK_M d\eta_\alpha
\end{equation} 
where $N_\alpha$ is given as

\begin{equation}
N_\alpha=\int \eta^4_\alpha({\bf r}) d{\bf r}.
\end{equation}  
Transforming $\delta$ function into an integral and carrying out the
Gaussian integral, we obtain, 

\begin{eqnarray}
&&P(\Delta)= N \int 
D\eta_\alpha({\bf r})
\frac{{2\Delta N_\alpha}}{\Delta^2_0}
\frac{1}{\sqrt{2\pi\sigma_\alpha(\{\eta({\bf r})\})}}
\exp[- \frac{(E_\alpha(\{\eta({\bf r})\})  
+ N_\alpha (\frac{\Delta}{\Delta_0})^2)^2 }  
{{2\sigma_\alpha(\{\eta({\bf r})\})}}]
\end{eqnarray}
The saddle point equation of Eq.59 is similar to Eq.43 except
there is an additional nonlinear term proportional to $N_\alpha$.
As we will see that the typical $\Delta$ in optimal droplet is 
much smaller than $\Delta_0\sqrt{H-H_c^0/H_c^0}$, 
this new term is much smaller than the linear term and can be
neglected as far as the spatial dependence is concerned. 
We can use the saddle point solution obtained in Eq.43 to evaluate Eq.59,

\begin{equation}
P(\Delta^2)
=\frac{2 N_s \Delta}{\Delta_0^2}
\frac{1}{\sqrt{2\pi\sigma_s}}\exp(-\frac{I_{s}^2}{{2\sigma_{s}}})
\frac{det \Gamma({\bf r}, {\bf r'})}{det <{\cal O}({\bf r}, {\bf r'})>}
\end{equation}  
where

\begin{equation}
I_{s}=E_{s} + N_s\frac{\Delta^2}{\Delta_0^2}.
\end{equation} 
$N_s$ is the corresponding value of $N_\alpha$ evaluated 
at $\eta_\alpha({\bf r})=\eta_s({\bf r})$.
Substituting the results in Eqs.51,52 into Eq.60, we obtain the
conditional distribution function of $\Delta$

\begin{equation} 
P_c(\Delta^2)=\frac{2CG\Delta}{\Delta^2_0}
\exp(-C^2 G^2\frac{\Delta^2}{\Delta^2_0}).
\end{equation}

\section{Exchange interaction between the droplets}

The coupling between different droplets deserves special attention.
Though the coupling between droplets 
does not affect the probability of finding
one droplet, it determines the global phase rigidity.
The typical distance $L_d$ is of order 
\begin{equation}
L_f {\rm erfc}^{-1/2}[\frac{BG}{A} (\frac{H-H_c^0}{H_c^0})^{1/2}]
\end{equation}
following Eq.56.
It is important that as long as $H-H_c/ H_c \ll 1$, 
$L_d$ is much less than $\xi_c=
l \exp(-G^2)$, the localization length in the presence of 
a parallel magnetic field; the weak localization effect in this
case is small as far as the superconductivity is concerned. 
The typical coupling between $\alpha$ and $\beta$ droplet
is determined by $\delta {\cal O}$.
Taking into account Eq.51, 
in the limit $L_d \gg L_f$
we obtain the variance of the coupling 

\begin{equation}
\sqrt{<\left(\nu_0 \Delta_\alpha \Delta_\beta 
\int d{\bf r} d{\bf r}' \delta K^0({\bf r},
{\bf r}')\eta_s({\bf r}-{\bf r}_\alpha)
\eta_s({\bf r}' - {\bf r}_\beta)\right)^2>} \propto 
\frac{\Delta_0}{G^2}
\left(\frac{L_f}{L_d}\right)^2.
\end{equation}
(The coupling depends on $\Delta_0$ in this case because the 
spectrum in the superconducting droplet is gapless.)
To get this result, we take into account that the size of the droplet
is $L_f$, typical $\Delta_\alpha$ is given by Eq.62 
and $|{\bf r_\alpha} -{\bf r_\beta}| \sim L_d$. 
One the other hand, the average ${\cal O}$, as shown in Eq.19
is proportional to $\delta({\bf r}-{\bf r}')$.
The average coupling
is proportional to the overlap integral of the
wave functions of two droplets 
\begin{equation}
\int d{\bf r} \eta_s({\bf r}-{\bf r}_\alpha)
\eta_s({\bf r} -{\bf r}_\beta) \propto 
\exp(-L_d/L_f).
\end{equation}
The variance of the coupling evaluated in Eq.64
is much larger than the average coupling 
in the limit $L_d \gg L_f$.  
The distribution function
of the coupling between different pairing states is symmetric with 
respect to zero 
and the sign of the coupling between different 
mesoscopic pairing states (droplets) is random.
This suggests that the 
ground state of these coupled mesoscopic
pairing states will exhibit glassy behavior in this limit.

The existence of random Josephson coupling in the 
presence of a parallel magnetic field is a consequence
of the Pauli spin polarization. This phenomenon exists 
even without {\em spin orbit scattering}.
Consider
for example a granular superconductor,
with superconducting grains embedded inside a
{\em noninteracting} disordered metal coupled with each other
via Josephson coupling. 
The sign of the Josephson coupling
is determined by the total phase of the time reversal
pairs. In the pure limit, though the sign of the 
wave function of each electron oscillates with 
a period of the Fermi wave length, the total
phase of $(\bf p, -\bf p)$ pair
is zero because of the exact cancellations of the 
phases of each electron inside the pair. 
Therefore there is no sign oscillation
for Josephson couplings.
In the dirty case $\bf p$ is not a good quantum number.
However the sign of the coupling does not oscillate as a
function of spatial coordinate because of the time reversal
symmetry. As a result, even 
when the distance between the grains is much larger than
the mean free path, 
the sign of the coupling is positive defined\cite{Coupling}. 
This is in contrast to 
RKKY exchange interaction between nuclear spins.
RKKY coupling exhibits
Friedel oscillations with the period of the Fermi wave length
in the pure case;  
in the presence of impurity scattering, the phase of Friedel
oscillations of electron wave functions becomes
random.

In the presence of a parallel magnetic field,
the electrons inside the normal metal become polarized. 
In this case, the electron with spin up
has a different kinetic energy as the electron with spin down on the Fermi
surface because of the Pauli spin polarization. As a result,
the phase of the electron with the spin up does not cancel with  
that of the spin down one in the presence of Zeeman splitting, and 
the total phase $\phi$ is equal to
$\int_{\cal C}  
{\bf p}_{up} \cdot d{\bf r} +\int_{\cal C} {\bf p}_{down} \cdot d{\bf r}
=\int_{\cal C} d{\bf r} {\mu_B H}/{v_F}$.
Integral is carried out along trajectory ${\cal C}$ along which
electron pairs travel.
In the pure limit, $\phi \approx L \mu_B H /v_F$
with $L$ the distance between two grains.
The pairing wave function oscillates and develops nodes
in its spatial dependence

\begin{eqnarray}
2\pi kT \sum_\omega \sigma^{\alpha-\alpha}_{y}\sigma^{\beta-\beta}_y
G^{\alpha\beta}_{\omega}({\bf r}, {\bf r}')
G^{-\alpha-\beta}_{-\omega}({\bf r}, {\bf r}')
=\frac{\nu_0}{|{\bf r}-{\bf r}'|^2} 
\cos(\frac{|{\bf r}-{\bf r'}|\mu_BH}{v_F}-\frac{\pi}{4})
\end{eqnarray}

This leads to the sign oscillations of 
the Josephson coupling
with a period $v_F/\mu_B H$, which is much longer than the 
Fermi wave length.
The positions of these nodes in the spatial
dependence of the coupling 
can be shifted in random
directions when impurities are present.
To estimate these random
phase shifts, consider disordered metals with short mean free path
and $L \gg l$.
The trajectory
of electron pairs is 
a diffusion path with typical length $L^2/l$.
In this case, $\phi \sim \mu_B H L^2/D$. 
When $L^2/l \gg v_F/\mu_B H$, 
$\phi$ is much larger than unity,
and the sign of the coupling becomes unpredictable
for different impurity configurations. 
In this limit,
the Josephson coupling averaged over impurity configurations
is exponentially small 
$\exp(-{\sqrt{2}L}/\sqrt{D/\mu_B H})$
while the typical amplitude of the coupling decays 
as ${L^{-2}}$. 
Therefore when the magnetic field increases, only the position
of the maximum of the distribution
function moves towards zero while the 
width of the distribution function barely changes.
This results in the superconducting glass state.
Note that in principle the charging effect inside the grain will also lead
to the 
superconducting glass phase as suggested in a recent experiment\cite{Wu95}.
However in the metallic limit when the tunneling conductance
between the grain and the normal metal is much larger than
$e^2/\hbar$, charging effect should be negligible
and only the mesoscopic mechanism discussed 
in this paper is relevant.

In this section,
we find that BCS order parameter is determined by 
mesoscopic fluctuations of physical quantities.
Short range mesoscopic fluctuations are responsible 
for the presence of optimal superconducting droplets 
while long wave length fluctuations lead to 
frustrations.  
Close or above the mean field critical points, the inhomogeneous 
superconducting states are described by a {\it nonlocal} Landau
-Ginsburg theory.

\section{Mesoscopic sample}

For a finite system of size $L$ 
there are, in principle, many critical fields $H_c^i$.
Linearizing Eq.33 with respect to $\Delta$,
neglecting the gradient term and using perturbation 
theory we have 
\begin{equation}
\delta H_{c}=H_{c}-H_{c}^{0}=H_c^0\frac{1}{L^2}
\int \delta K^{0}({\bf r},{\bf r'}, H^0_c + 
\delta H_c) d{\bf r}d{\bf r'}
\end{equation}
To derive Eq.67 we have taken into account: 1. The relative
amplitude of fluctuations of the critical field is smaller than its average
$\langle(\delta H_c)^{2}\rangle\ll (H_{c}^{0})^{2}$. 2. The sample size is
smaller than 
the coherence length $L < \xi(H)$ and $\Delta({\bf r})$ is spatially 
uniform. 
Eq.67 reflects the fact that the magnetic field acts on the system in two 
ways: a) It suppresses superconductivity, b)
It changes the mesoscopic fluctuations of parameters of the 
normal metal and the quantity $
\int \delta K^{0}({\bf r}, {\bf r}', H) d{\bf r'}d{\bf r}$ is a random
function of $\delta H_c$. Therefore, generally speaking, at a given $T$,
Eq.67 can have an infinite
number of solutions,
which means that the $H$-dependence of the critical temperature $T_{c}(H)$ 
exhibits reentrant superconductor-metal transitions as a function
of $H$. Qualitatively the picture of
the reentrant transitions is very similar to that which takes place in
the case of magnetic field induced orbital effects 
\cite{Spivak95}.
To characterize the random quantity
$\delta H_{c}$, we study the statistics of $s$, which is the right hand
side of Eq.67.
Straightforward calculation of its variance following Eq.15 yields
\begin{equation}
\sigma_0=<s^2>=\int d{\bf r}_1 d{\bf r}_2
{\cal C}({\bf r}_1, {\bf r}'_1; {\bf r}_2, {\bf r}'_2)
\end{equation}
Its distribution function in the Gaussian limit reads as
\begin{equation}
P^0(s)=\frac{1}{\sqrt{2\pi\sigma_0}}\exp(-\frac{s^2}{2\sigma_0}),
\sigma_0=A_2\frac{1}{G^2}\frac{\xi_0}{L^2}.
\end{equation}
Following Eq.67, the distribution of $H_c$ is 

\begin{equation}
P^0(H_c)=\int \delta({H_c- H^0_c}-{H^0_c}s(H_c)) P(s(H_c)) ds(H_c)
\nonumber \\
=\frac{1}{\sqrt{2\sigma_0}}\frac{1}{H_c}
\exp(-(\frac{H^0_c-H_c}{H^0_c})^2\frac{1}{2\sigma_0}). 
\end{equation}
In deriving the second line we use that

\begin{equation}
<\frac{\partial s}{\partial H} s>=0.
\end{equation}
It is obvious following Eq.70 that the variance of $H_c$ is
\begin{equation}
\frac{\langle\delta H_{c}^{2}\rangle}{(H^0_{c})^2}=
\frac{\xi_{0}^{2}}{L^2 G^{2}}.
\end{equation}
Eq.72 gives the interval of the magnetic field near
$H^{0}_{c}$ where the reentrance takes place with a probability of 
order of unity.
The probability for a sample in a superconducting state 
at $H$ can be estimated as 

\begin{equation}
{\cal P}_s(H)=\int^{+\infty}_{H} P^0(H)dH
=\frac{GL}{\xi_0\sqrt{2\pi A_2}} erfc(G\frac{L}{\xi_0\sqrt{2A_2}}
\frac{H-H_c}{H_c}).
\end{equation}

When spin-orbit 
scatterings are weak, $\Delta_{0}\tau_{so}\gg 1$, the 
conventional theory leads to the conclusion that 
the superconductor-normal metal transition is of first order 
with the critical magnetic field $H_{cc}$ \cite{Ch62,Clo62}. 
In this case the spin polarization in the superconducting phase is zero. 
The average spin polarization energy of a normal metal sample of
size $L$ and its mesoscopic
fluctuations are of order 
$d L^2 \nu_0(\mu_B H)^2$
and $d L \xi_0 \nu_0 (\mu_B H)^2G^{-1}$, respectively. 
As a result, a finite superconducting sample exhibits 
first order normal metal
-superconductor reentrant transitions in the interval of magnetic field
of order $\delta H=H_{cc}{\xi_0}/{GL}$ in the vicinity of the critical
field.

In the case of two dimensional superconducting film, the fluctuations of
both 
polarization energy of the normal metal and the condensation energy of
the superconducting phase should lead to a nonuniform state, 
similar to the case $\Delta_{0}\tau_{so}\ll 1$.
The theory of this phenomenon at $\Delta_{0}\tau_{so}\gg 1$ is, however,
more difficult.  In this case a domains of normal phase within a bulk 
superconductor (or a superconducting domain in normal metal) has the
surface energy of order of $d L_{D}\xi_{0}\nu_0\Delta_0^2$, where 
$L_{D}$ is the domain size. This energy is much larger than 
the typical energy associated with mesoscopic fluctuations in Eq.36.
Thus the probability of 
the occurrence of such domains is small even
at the critical point.  

It is worth emphasising that
qualitatively, the case 
$\Delta_{0}\tau_{so}\gg 1$ is not 
different from the case $\Delta_{0}\tau_{so}\ll 1$ for in both cases the
superconducting glass solutions survive at $T=0$ and $H>H^{0}_{c}$.    
Especially, for a quasi 1D thin stripe with the width $W\ll L_D$, the surface 
energy becomes independent of the size of the domain while the mesoscopic
fluctuations of spin polarization energy is proportional to $\sqrt{L_D}$.
This situation is similar to the strong spin orbit scattering limit
discussed before.

\section{conclusion}

We show the existence of a novel superconducting
glass phase in disordered thin films in Clogston limit. 
The statistics of mesoscopic pairing states in the superconducting
glass phase is universal and  
determined only by the sheet conductance. It is a direct consequency of
Wigner-Dyson statistics of single particle energy spectrum.

This allows us to distinguish the mechanism 
discussed in this paper and the effect of inhomogeneity of
impurity concentration, or classical pinning effect on vortex lattices
discussed in\cite{vin}. First of all, in the present case,
the magnetic field couples only with spins and
the wave functions are real(as far as the impurity averaged
condensate wave function is concerned); the time reversal symmetry is 
broken spontaneously. For classical pinning effects on vortex lattices, 
the time 
reversal symmetry is broken by the applied perpendicular magnetic field.
More over, fluctuations of local quantities like mean free path
can lead to inhomogeneous states but do not lead to spontaneous
time reversal symmetry breaking. 
The glass state discussed in this paper is due to random signs of long
range exchange interaction, which is purely of mesoscopic nature.
Finally, the response of the state discussed here is determined  
{\em universally} by Thouless energy of the size of the coherence length and 
the response of a pinned 
vortex glass depends very much on the range and strength of
the classical pinning potential.  For amorphous films 
where the impurity potential is perfectly screened and
in the absence of granularities,
the classical pinning effect is weak; the mesoscopic
effects dominate in this limit. Amorphous thin films like
Zn\cite{Okuma}, Mo-Ge\cite{Graybeal}, Pb\cite{Dynes}, 
In-InO\cite{Hebard}, 
Bi\cite{Goldman} 
have been subjects of extensive studies.

Though
the transport properties of such a superconducting glass
state are poorly understood,
it shares all the features a glass state has:
hysteresis, stretched relaxation time. 
Another experimental consequence of 
random sign of $N_{s}({\bf r})$ we like to mention is, following to 
Ref.\cite{KS}, at 
$H>H_{SG}$ and at a finite temperature the system exhibits
the negative magnetoresistance with respect to the component of the
magnetic field perpendicular to the film.

For a finite sample,
when gate voltages are applied, the mesoscopic 
fluctuations in Eq.67 start to oscillate. This causes 
reentrant superconductor-metal phase transitions, similar
to the magnetic field induced reentrant transitions. Such 
transitions should manifest themselves in the gate voltage finger
print experiment: the conductance as a function of gate voltages 
exhibits sample specific fluctuations, 
with amplitude equal to the normal sample conductance.  
The conductance fluctuation can much exceed the value of UCF
due to the attractive interaction!\cite{zhou98}.

The other possibility to study the mesoscopic
superconductor is to bring the superconducting state adiabatically
along a closed trajectory in a parameter space
via applying gate voltages(for thin films like Bi, the
chemical potential can be varied by 20 percent.).
Adiabatic charge transport across a boundary of the system
per period in the presence of 
periodically changing external perturbations is 
connected with a geometric phase, as first pointed out
by D. Thouless\cite{Thouless}. 
Recently, this idea was applied to normal metal mesoscopic systems where 
quantum chaos is fully developed; the charge transport is determined by 
the amount of "flux" of a topological field which
threads the area enclosed by a closed trajectory in the parameter space 
\cite{Brouwer,zhou99}.
In a normal metal mesoscopic sample, such a topological 
field was shown to be determined by the sensitivity of the
quantum chaos to external perturbations, which is a random
quantity. 
In the case of superconductors,
the geometric phase will be determined by
the compressibility of the superfluid density
because excess electronic density  created by gate voltages can be carried 
away only by coherent motions of the condensate. The mesoscopic 
fluctuations of superfluid density are "more compressible"
than the electronic density itself, i.e.

\begin{equation}
\frac{\partial \delta {\cal N}_{ij}}{\partial \mu}
\gg \frac{\partial {N_s}}{\partial \mu}.
\end{equation}
$\mu$ is the chemical potential. 
The existence of mesoscopic fluctuations of condensate
wave function or superfluid density manifests itself in a geometric 
phase. This problem will be addressed elsewhere.

The question whether or not the quantum fluctuations of the phase of the 
order parameter
destroy the superconducting glass state at $T=0$ and large $H$
is still open.
We would like to mention here, that a similar question was 
addressed in many papers in the context of the 
disorder driven superconductor-insulator transition \cite{FishL,KCG,KE} 
and metallic spin glasses with 
dissipation \cite{Sa}. 

Strictly speaking, 
at arbitrary finite temperatures $T>0$, 
a superconducting
film doesn't possess a superconducting phase
rigidity.
In two dimensional 
case, due to screening, the interaction energy
between vortices decays as a power law rather than logarithmically.
This leads to a finite concentration of   
unbounded vortices with the correlation function 
of the phase of the order parameter decaying exponentially at large
distances \cite{tho}.

However, in real experiment situations,
the London penetration length can be comparable or longer
than the sample size. Furthermore, the exchange interaction
decays as a power law function, $1/r^2$, as shown in Eqs.15,28.
The typical energy of a domain of size $L$
diverges logarithmically as $L \rightarrow \infty$, 
suggesting that there could be
a finite temperature phase transition between the superconducting 
glass phase and a normal phase.
In Fig.1, we plot a dashed line which separates these two phases,
the existence of which needs further investigation. 
On the other hand the two dimensional $XY$ model with 
{\em short range} random exchange 
interaction is known not to exhibit a phase transition between the 
paramagnetic and the spin-glass phases \cite{bray}.

The frustration which leads to the novel superconducting glass
phase is due to the existence of nodes in the condensate wave function
when spins are polarized, as emphasized in section 5. 
For a d-wave superconductor, nodes exist even in the 
absence of an external magnetic field;  
naturally one can ask whether a d-wave superconductor can be
free of frustration when disordered. We are not aware of
work on this subject and believe the answer to this question is
also critical to the understanding of the density of states at the Fermi 
surface in a disordered d-wave superconductor.

The author acknowledges useful discussions with B. Altshuler, 
C. Biagini, D. Huse, S. Kivelson, I. Smolyarenko, 
B. Spivak, N. Wingreen.
This work is supported by ARO under DAAG 55-98-1-0270.
He is also grateful to NECI, Princeton for its 
hospitality. 

\newpage

\section{Appendix}

\subsection{Fluctuations of spin polarization energy} 
Diagrams in Fig.2d yield

\begin{eqnarray}
\sqrt{\langle(\delta E_{p})^{2}\rangle}=
(2\pi T)^2\sum [
\sigma_z^{\beta_1 \beta_1} C_{\epsilon_1+\epsilon_2}^{\beta_1 \alpha_2}
\sigma_z^{-\alpha_2- \alpha_2} C_{\epsilon_1+\epsilon_2}^{\alpha_2 \beta_2}
\sigma_z^{\beta_2 \beta_2} C_{\epsilon_1 +\epsilon_2}^{\beta_2 \alpha_1}
\sigma_z^{-\alpha_1 -\alpha_1} C_{\epsilon_1+\epsilon_2}^{\alpha_1 \beta_1} 
\nonumber \\
+\sigma_z^{\beta_1 \beta_1} D_{\epsilon_1+\epsilon_2}^{\beta_1 \alpha_2}
\sigma_z^{-\alpha_2 -\alpha_2} D_{\epsilon_1+\epsilon_2}^{\alpha_2 \beta_2}
\sigma_z^{\beta_2 \beta_2} D_{\epsilon_1 +\epsilon_2}^{\beta_2 \alpha_1}
\sigma_z^{-\alpha_1 -\alpha_1} 
D_{\epsilon_1+\epsilon_2}^{\alpha_1 \beta_1}]. 
\end{eqnarray}
$\sigma_z^{\alpha\alpha}=\alpha$, $\alpha=\pm 1$ is for spin index;
$\epsilon=(2n+1)\pi kT$ is Matsubara frequency.
$\sum$ represents summations
over Matsubara frequency, momentum and spin index.

Following Dyson equation in Fig.2c,

\begin{eqnarray}
&&C_{\omega}^{\alpha\beta}({\bf q}^2)
=\frac{1}{2}[\frac{\alpha\beta}{|\omega|+D{\bf q}^2+\tau_{so}(\mu_B H)^2}
+\frac{1}{|\omega|+D{\bf q}^2+\tau^{-1}_{so}}]
\nonumber \\
&&D_{\omega}^{\alpha\beta}({\bf q}^2)
=\frac{1}{2}[\frac{\alpha\beta}{|\omega|+D{\bf q}^2}
+\frac{1}{|\omega|+D{\bf q}^2+\tau^{-1}_{so}}]
\end{eqnarray}
The sensitivity to the change of pair potential $\delta \Delta$ is 
given by Eq.75 with
$C, D$ replaced via $\delta C, \delta D$ 

\begin{eqnarray}
&& \delta D^{\alpha\beta}_{\epsilon_1+\epsilon_2}
=(\delta\Delta)^4\sum (|\epsilon_1+\epsilon_2|+2 D{\bf q}{^2})^2
\sigma_y^{\gamma_3 -\gamma_3} 
\sigma_y^{\gamma_4 -\gamma_4} 
\sigma_y^{\gamma_5 -\gamma_5} 
\sigma_y^{\gamma_6 -\gamma_6} 
\nonumber \\
&&D_{\epsilon_1+\epsilon_2}^{\alpha\gamma_1}({\bf q}^2)
C_{2\epsilon_1}^{\gamma_1\gamma_3}(0) 
C_{2\epsilon_1}^{\gamma_2\gamma_4}(0)
D_{\epsilon_1+\epsilon_2}^{\gamma_1 \gamma_2}({\bf q}^2) 
C_{2\epsilon_2}^{\gamma_1\gamma_5}(0) 
C_{2\epsilon_2}^{\gamma_2\gamma_6}(0) 
D_{\epsilon_1+\epsilon_2}^{\gamma_2 \beta}({\bf q}^2) 
\nonumber \\
&&\delta C^{\alpha\beta}_{\epsilon_1+\epsilon_2}
=(\delta\Delta)^4\sum (|\epsilon_1+\epsilon_2|+2 D{\bf q}{^2})^2
\sigma_y^{\gamma_3 -\gamma_3} 
\sigma_y^{\gamma_4 -\gamma_4} 
\sigma_y^{\gamma_5 -\gamma_5} 
\sigma_y^{\gamma_6 -\gamma_6} 
\nonumber \\
&&C_{\epsilon_1+\epsilon_2}^{\alpha\gamma_1}({\bf q}^2)
C_{2\epsilon_1}^{\gamma_1\gamma_3}(0) 
C_{2\epsilon_1}^{\gamma_2\gamma_4}(0)
C_{\epsilon_1+\epsilon_2}^{\gamma_1 \gamma_2}({\bf q}^2) 
C_{2\epsilon_2}^{\gamma_1\gamma_5}(0) 
C_{2\epsilon_2}^{\gamma_2\gamma_6}(0) 
C_{\epsilon_1+\epsilon_2}^{\gamma_2 \beta}({\bf q}^2), 
\end{eqnarray}
following Fig.2e.

\subsection{Fluctuations of superfluid density}

The correlation function of the mesoscopic fluctuations of the superfluid
density consists of two terms. First term is given in Fig.2g

\begin{eqnarray}
&&\delta_{ij}\delta_{i'j'}
\delta({\bf r}_1 -{\bf r}'_1)
\delta({\bf r}_2 -{\bf r}'_2) N^2_s(H) 
\frac{\langle \delta \Delta({\bf r}_1)
\delta \Delta({\bf r_2})\rangle^2}{\langle\Delta(H)\rangle^{4}}
\times
\nonumber \\
&&(2\pi k T)^2\Delta^2_0 
\sum_{\epsilon_1\epsilon_2}
\sigma_y^{\alpha_1 -\alpha_1}
\sigma_y^{\alpha_2 -\alpha_2} 
\sigma_y^{\beta_1 -\beta_1}
\sigma_y^{\beta_2 -\beta_2} 
C_{2\epsilon_1}^{\alpha_1 \gamma_1}(0)
C_{2\epsilon_1}^{\gamma_1 \alpha_2}(0) 
C_{2\epsilon_2}^{\beta_1 \gamma_2}(0)
C_{2\epsilon_2}^{\gamma_2 \beta_2}(0), 
\nonumber \\
\end{eqnarray}
while the second part of the contribution in Fig.2h 

\begin{eqnarray}
&&\{\delta_{ii'}\delta_{jj'}\delta({\bf r}_1 -{\bf r}_2)
\delta({\bf r}'_1 -{\bf r}'_2)
+\delta_{ij'}\delta_{i'j}\delta({\bf r}_1 -{\bf r}'_2)
\delta({\bf r}_2 -{\bf r}'_1)\} N_s^2(H)\times
\nonumber \\
&&(2\pi kT)^2\frac{\Delta_0^2}{\nu_0^2}\sum
\sigma_y^{\alpha_1 -\alpha_1}\sigma_y^{\alpha_2 -\alpha_2}
\sigma_y^{\beta_1 -\beta_1}\sigma_y^{\beta_2 -\beta_2}
(\epsilon_1 + \epsilon_2)^2
\nonumber \\
&&[C^{\alpha_1\gamma_1}_{2\epsilon_1}(0)
C^{\beta_1-\gamma_1}_{2\epsilon_2}(0)
C^{\gamma_2\alpha_2}_{2\epsilon_1}(0)
C^{-\gamma_2\beta_2}_{2\epsilon_2}(0)
C^{\gamma_1 \gamma_3}_{\epsilon_1+\epsilon_2}({\bf q}^2)
C^{\gamma_3 \gamma_2}_{\epsilon_1+\epsilon_2}(({\bf q'})^2)
C^{-\gamma_1 \gamma_4}_{\epsilon_1+\epsilon_2}(({\bf q})^2)
C^{\gamma_4 -\gamma_2}_{\epsilon_1+\epsilon_2}({\bf q'}^2)
+ \nonumber \\
&&C^{\alpha_1\gamma_1}_{2\epsilon_1}(0)
C^{\beta_1\gamma_1}_{2\epsilon_2}(0)
C^{\gamma_2\alpha_2}_{2\epsilon_1}(0)
C^{\gamma_2\beta_2}_{2\epsilon_2}(0)
D^{\gamma_1 \gamma_3}_{\epsilon_1+\epsilon_2}({\bf q}^2)
D^{\gamma_3 \gamma_2}_{\epsilon_1+\epsilon_2}(({\bf q'})^2)
D^{-\gamma_1 \gamma_4}_{\epsilon_1+\epsilon_2}(({\bf q})^2)
D^{\gamma_4 -\gamma_2}_{\epsilon_1+\epsilon_2}({\bf q'}^2)].
\nonumber \\
\end{eqnarray}

\subsection{$\rho_\alpha$ when $\delta {\cal O}$ is non-Gaussian}

In general, statistics of $\delta{\cal O}$ is characterized  by the 
following correlators, 

\begin{equation}
<\Pi^{2m}_{1}\delta {\cal O}({\bf r}_i, {\bf r}'_i)>
={\cal C}_{2m}({\bf r}_i, {\bf r}'_i|i=1,...,2m), 
\end{equation}
Eq. 43 then is transformed into
$P_l(\{\eta_\alpha\}|\alpha=1,...,l)
=\Pi_{\alpha}\rho_\alpha$, 
where

\begin{eqnarray}
&&\rho_\alpha=N \int_{-\infty}^{0} dg_\alpha 
\int_{-\infty}^{+\infty}dh_\alpha 
D\eta_\alpha({\bf r})
\exp(ih_\alpha (L_\alpha-g_\alpha))
\nonumber \\ &&\sum_m\frac{h^{2m}}{2m!}
\int \Pi_{k=1,...,2m}\eta_\alpha({\bf r}_k) 
\eta_\alpha({\bf r}'_k)d{\bf r}_k d{\bf r}'_k 
{{\cal C}}_{2m}(\{{\bf r}_k, {\bf r}'_k\}|k=1,...,2m) \nonumber \\
&&=\lim_{b\rightarrow 0} \sum_{2m} \frac{1}{2m!}
\frac{\partial^{2m}}{\partial b^2} Erf(L_\alpha, b^2) 
\int \Pi_{k=1,...,2m}\eta_\alpha({\bf r}_k)\eta_\alpha({\bf r}'_k)d{\bf
r}_k d{\bf r}'_k 
{{\cal C}}_{2m}(\{{\bf r}_k, {\bf r}'_k\}|k=1,...,2m). 
\end{eqnarray}
The main contribution is from the saddle point where

\begin{eqnarray}
&&\frac{\delta \rho(\{\eta({\bf r})\})}{\delta\eta({\bf r})}=
O_{LG} \eta({\bf r}) + V(\{\eta({\bf r})\})=0
\nonumber \\
&&V(\{\eta({\bf r})\})=\eta^{-1}_\alpha({\bf r})(\frac{\partial
\rho_\alpha}{\partial L_\alpha})^{-1}\nonumber\\
&&\lim_{b\rightarrow 0} \sum_{2m} \frac{1}{2m!}
\frac{\partial^{2m}}{\partial b^2} Erf(L_\alpha, b^2) 
\int \Pi_{k=1,...,2m}\eta_\alpha({\bf r}_k)\eta_\alpha({\bf r}'_k)d{\bf
r}_k d{\bf r}'_k 
\delta({\bf r}-{\bf r}_k){\cal C}_{2m}(\{{\bf r}_k, {\bf r}'_k\}|k=1,...,2m).
\nonumber \\
\end{eqnarray}
In the Gaussian limit, 
Eq.82 yields Eq.43.

\newpage
\begin{figure}
\begin{center}
\leavevmode
\epsfbox{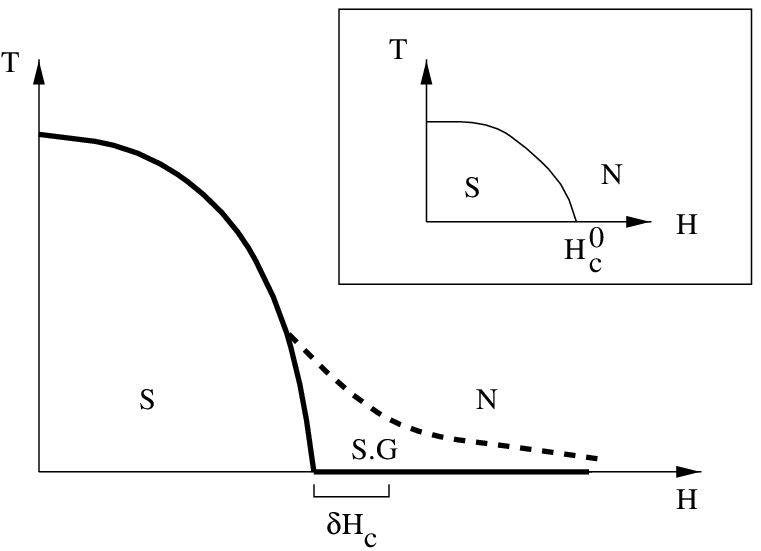}

\caption{
Suggested phase diagram in Clogston limit when 
$\Delta_0\tau_{s0} \ll 1$; 
Insert is the phase diagram of Chandrasekhar-Clogston theory.
A superconducting glass phase  
appears at $H_{SG}=H^0_c-\delta H_c$ with 
$\delta H_{c} \sim H^0_{c}/G^2 \ln G$. The thick line along $T=0$ axis
represents the superconducting glass phase discussed in the paper; the dashed 
line stands for a possible finite temperature phase transition
between the superconducting glass phase and a normal phase.}
\end{center}

\begin{center}
\epsfbox{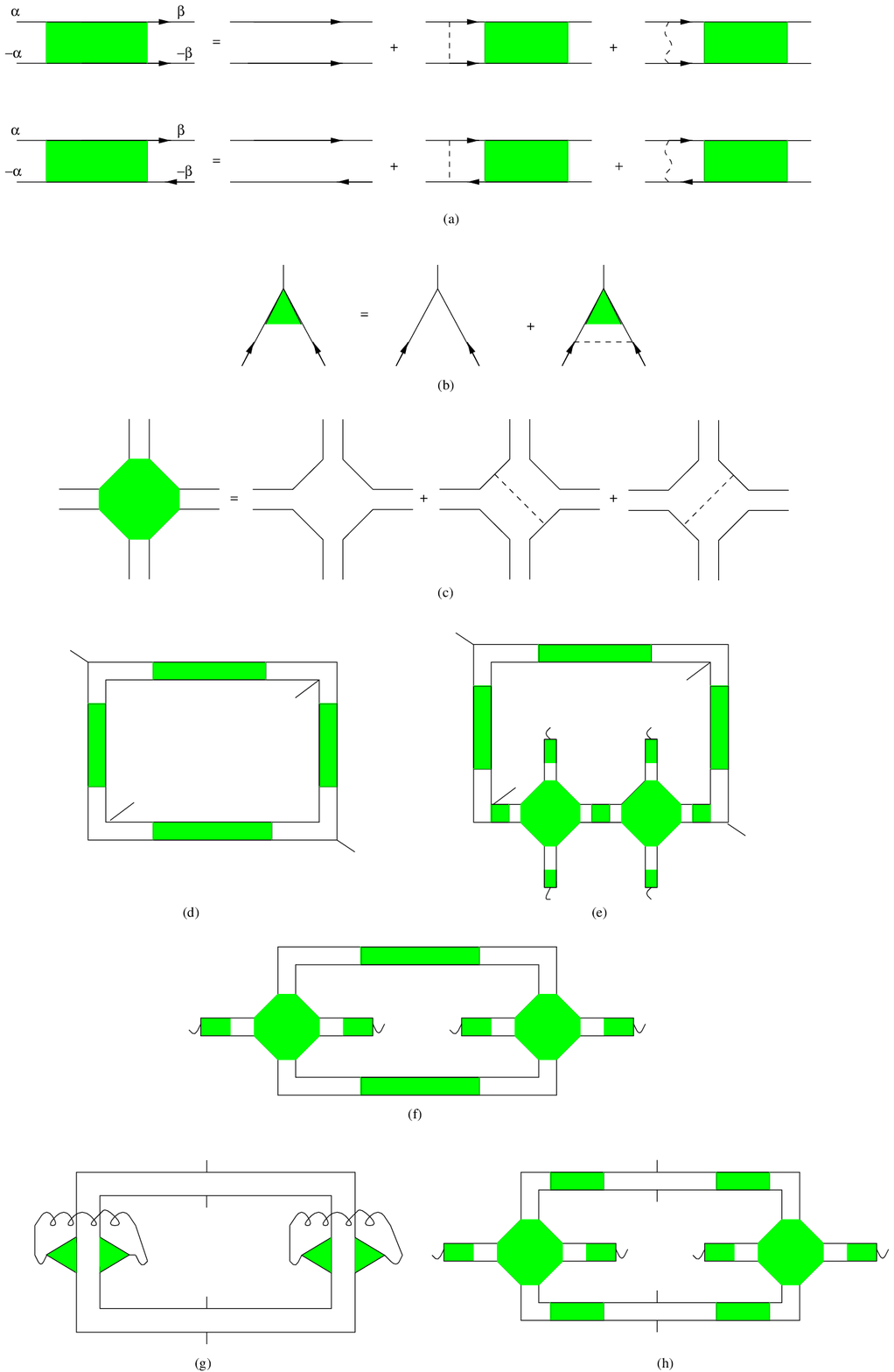}
\caption{
a). Dyson equation for diffusons and Cooperons.
Solid lines correspond to
electron Green functions in a metal; 
Dashed lines are for impurity scatterings preserving the spin
while the wavy ones for spin orbit scatterings.
b).Vertex correction of pair potentials.
c). Hikami Box.
d),e). Diagrams for the fluctuations of spin polarization energy. 
f) Diagrams representing the correlation function $\langle\delta
K^{0}({\bf r}_{1},{\bf r}_{2})\delta
K^{0}({\bf r}_{3},{\bf r}_{4})\rangle$. 
g),h). Diagrams representing the correlation function of
supercurrent densities
$\langle{\bf J}_{s}({\bf r}){\bf J}_{s}({\bf r'})\rangle$.
Solid wavy lines in g). represent the
correlation function $\langle\delta \Delta({\bf r}_1) 
\delta \Delta({\bf r}_2)\rangle$ given in Eq. 17.
}

\end{center}

\end{figure}

\begin{thebibliography}{99}
\bibitem{Wu95}W. Wu, P. W. Adams, Phys. Rev. Lett.{\bf 74}, 610(1995).
\bibitem{Abr}A. A. Abrikosov, {\it 
Fundamentals of The Theory of Metals}, North-Holland, 1988.
\bibitem{Clo62}A. M. Clogston, Phys. Rev. Lett.{\bf 9}, 266(1962).
\bibitem{Ch62}B. S. Chandrasekhar, Appl. Phys. Lett. {\bf 1}, 7(1962).
\bibitem{Lee}P. A. Lee, D. Stone, Phys. Rev. Lett.{\bf 55}, 1622(1985).
\bibitem{Altshuler85}B. Altshuler, Pisma Zh. Eksp. Teor. Fiz. 
{\bf 41}, 530(1985)[JETP Lett. {\bf 41}, 648(1985)]
\bibitem{Altshuler86}B. Altshuler, B. I. Shklovskii,
Zh. Eksp. Teor. Fiz. {\bf 91}, 220(1986)  
[Sov. Phys. JETP. {\bf 64}, 127(1986)]
\bibitem{Finkelshtein}A. M. Finkelshtein, Pisma Zh. Eksp. Teor. Fiz.
{\bf 45}, 37(1987)[JETP Lett. {\bf 45}, 46(1987)] and references
therein.
\bibitem{Edward}S. F. Edwards, P. W. Anderson, J. Phys. {\bf F 5},
965(1975).
\bibitem{Alt}B. Altshuler B. Spivak, 
Zh. Eksp. Teor. Fiz. {\bf 92}, 609(1987)
[Sov. Phys. JETP {\bf 65},343(1987)].
\bibitem{ZyS}B. Spivak, A. Zyuzin,
Pisma Zh. Eksp. Teor. Fiz. {\bf 47}, 221(1988) 
[Sov. Phys. JETP Lett. {\bf 47}, 267(1988)].
\bibitem{Spivak91}B. Spivak, A. Zyuzin, {\it
Mesoscopic fluctuations of Current Density in Disordered
Conductors}, Mesoscopic Phenomena in Solids edited by
B. Altshuler, P. Lee and R. Webb, Elsevier Science Publishers B. V.,
1991.
\bibitem{bin}C. W. J. Beenakker,
Phys. Rev. Lett. {\bf 67}, 3836(1991).
\bibitem{Spivak95}B. Spivak, F. Zhou, Phys. Rev. Lett. {\bf 74}, 2800(1995).
\bibitem{Ar}A. Aronov, B. Spivak, Fiz. Tverd. Tela {\bf 17}, 2806 (1975)
[ Sov. Phys. Solid State  17,1874 (1975)].
.\bibitem{Bul}L. Bulaevski, V. Kuzzi, 
A. Sobianin, 
Pisma Zh. Eksp. Teor. Fiz. {\bf 25}, 314(1977) 
[JETP Lett. {\bf 25}, 7(1977)].
\bibitem{Ki}B. Spivak, S. Kivelson, Phys. Rev. {\bf B 43}, 3740(1991).
\bibitem{KS}S. Kivelson, B. Spivak, Phys. Rev. {\bf B 45},10492(1990).
\bibitem{zhou98b}F. Zhou, B. Spivak, Phys. Rev. Lett. {\bf 80}, 5647(1998).
\bibitem{Werthamer66}N. R. Werthamer, E. Helfand, P. C. Hohenberg,
Phys. Rev. {\bf B 147}, 295(1966).
\bibitem{Maki}K. Maki, Phys. Rev. {\bf B 148}, 362(1966).
\bibitem{Ferrell}R. A. Ferrell, Phys. Rev. Lett. {\bf 3}, 262(1959). 
\bibitem{An}P. W. Anderson, Phys. Rev. Lett. {\bf 3}, 325(1959). 
\bibitem{Ab62}A. A. Abrikosov, L. P. Gorkov, Sov. Phys. JETP {\bf 15},
752(1962). 
\bibitem{Abrikosov62}A. A. Abrikosov, L. P. Gorkov, I. E. Dzyaloshinski,
{\it Methods of Quantum Field Theory in Statistical Physics},
Dover, 1975.
\bibitem{Kagan}
We want to thank Professor Y. Kagan for raising this issue.
\bibitem{Yo}H. Yoshika, Jour. Phys. Soc. Japan {\bf 63}, 405(1994). 
\bibitem{vil}J. Villain, Jour. Phys. {\bf C 10}, 4793(1977).
\bibitem{Lub}Sajeev John, T. C. Lubensky, Phys. Rev. {\bf B 34}, 4815(1986).
\bibitem{vin}V. Vinokur, L. Ioffe, A. Larkin, M. Feigelman, Sov. Phys.
JETP {\bf 66},198(1987).
\bibitem{FishG}M. Fisher, Phys. Rev. Lett. {\bf 62}, 1415(1989).
\bibitem{Fulde64}P. Fulde, P. A. Ferrell, Phys. Rev. {\bf B 135}, A550(1964).
\bibitem{Larkin64}A. I. Larkin, Yu. N. Ovchinnikov, Zh. Eksp.
Teor. Fiz. {\bf 47}, 1136(1964) [Sov. Phys. JETP {\bf 20}, 762(1965)]
\bibitem{asl}  L.G. Aslamazov, Zh. Eksp. Teor. Fiz. {\bf 55}, 1477-1482 (1968)
[Sov. Phys. JETP {\bf 28}, 773(1969)].
\bibitem{zhou98a}F. Zhou, C. Biagini, Phys. Rev. Lett. {\bf 81}, 4188(1998).
\bibitem{Lifshitz}I. M. Lifshitz, Adv. Phys. {\bf 13}, 483(1964).
\bibitem{Halperin}B. I. Halperin, M. Lax, Phys. Rev.{\bf 148},
722(1966).
\bibitem{Zittartz}J. Zittartz, J. S. Langer,  Phys. Rev.{\bf 148},
741(1966).
\bibitem{Igor}I. E. Smolyarenko, B. L. Altshuler, Phys. Rev. {\bf B 55},
10451(1977).
\bibitem{AKL}B. Altshuler, B. Kravtsov, I. Lerner, Sov. Phys. JETP {\bf
64},1352(1986).
\bibitem{KM}B. A. Muzykantskii, D. E. Khmelnitskii, Phys. Rev. {\bf B 51}, 
5480(1995). 
\bibitem{EF}K. Efetov, V. Falko, Phys. Rev. {\bf B 52}, 17413(1995).
\bibitem{Coupling}Strictly speaking, this statement is correct only 
when the bulk of the distribution function is concerned.
In principal, the time reversal symmetry 
doesn't exclude the possibility to have 
negative Josephson couplings. Furthermore,
in the presence of prelocalized states,
the correlational effect discussed in 
Ref.15,16
can also lead to a negative
superfluid density. However, the probability to find
such events is exponential small.
\bibitem{zhou98}F. Zhou, C. Biagini, Phys. Rev. Lett. {\bf 81}, 4724(1998).
\bibitem{Okuma}S. Okuma, F. Komori, Y. Ootuka, S. Kobayashi, J. Phys.
Soc. Jpn. {\bf 52}, 2639(1983).
\bibitem{Graybeal}J. M. Graybeal, M. R. Beasley, Phys. Rev. {\bf B 29},
4167(1984).
\bibitem{Dynes}R. C. Dynes, A. E. White, J. M. Graybeal, J. P. Garno,
Phys. Rev. Lett. {\bf 57}, 2195(1986).
\bibitem{Hebard}A. F. Hebard, M. A. Paalanen, Phys. Rev. Lett. {\bf 65},
927(1990).
\bibitem{Goldman}Y. Liu, K. A. McGreer, B. Nease, D. B.
Haviland, G. Martinez, J. W. Halley and 
and A. M. Goldman, Phys. Rev. Lett.{\bf 67}, 2068(1991).
\bibitem{Thouless}D. Thouless, Phys. Rev. {\bf B27}, 6083(1983).
\bibitem{Brouwer}P. Brouwer, Phys. Rev. {\bf B58}, 10135(1998).
\bibitem{zhou99}F. Zhou, B. Spivak, B. Altshuler, Phys. Rev. Lett. 
{\bf 82}, 608(1999).
\bibitem{FishL}M. Fisher, Phys. Rev. Lett. {\bf 57}, 885(1986).
\bibitem{KCG}S. Chakravarty, S. Kivelson, G. Zimanyi, B. Halperin,
Phys. Rev. {\bf B 35}, 7256(1987). 
\bibitem{KE}V.J. Emery, S.A. Kivelson, Phys. Rev. Lett. {\bf 74}, 3253
(1995). 
\bibitem{Sa}S.Sachdev, cond-mat/{\bf 9705074}.
\bibitem{tho}J. M. Kosterlitz, D. J. Thouless, J. Phys. {\bf C 6}, 1181(1973).
\bibitem{bray}B. W. Morris, S. G. Colborne, M. A. Moore, A. J. Bray,
J. Canisius, J. Phys. {\bf C 19}, 1157(1986).
\end{thebibliography}
\end{document}